\documentstyle[12pt,epsf,rotate]{article}

\def\be{\begin{equation}}
\def\ee{\end{equation}}

\def\feh{{\rm Fe/H}}

\def\mcommand#1{\ifmmode #1\else $#1$\fi}
\def\unit#1{\mcommand{\;{\rm #1}}}
\def\Msun{\mcommand{M_\odot}}
\def\sci#1{\mcommand{\times 10^{#1}}}
\def\dex{\unit{dex}}
\def\yr{\unit{yr}}
\def\Gyr{\unit{Gyr}}
\def\kmps{\unit{km/s}}

\def\element#1#2{\mcommand{{\vphantom{\rm #2}^{#1}{\rm #2}}}}
\def\D{{\rm D}}
\def\He#1{\element{#1}{He}}
\def\Li#1{\element{#1}{Li}}
\def\C#1{\element{#1}{C}}
\def\N#1{\element{#1}{N}}
\def\O#1{\element{#1}{O}}
\def\Ne#1{\element{#1}{Ne}}
\def\Si#1{\element{#1}{Si}}
\def\S#1{\element{#1}{S}}
\def\Fe#1{\element{#1}{Fe}}

\def\etal{{\em et al.}}
\def\hii{H~{\sc ii}}

\def\ltsim{\mathrel{\mathpalette\fun <}}
\def\gtsim{\mathrel{\mathpalette\fun >}}
\def\fun#1#2{\lower3.6pt\vbox{\baselineskip0pt\lineskip.9pt
  \ialign{$\mathsurround=0pt#1\hfil##\hfil$\crcr#2\crcr\sim\crcr}}}

\newdimen\boxdim
\setbox0=\hbox{$\Box$}
\boxdim=\ht0
\def\solidBox{\vrule height\boxdim width\boxdim depth0pt}

\def\chemevil{chemical evolution}

\makeatletter

\def\title#1{\gdef\@title{\uppercase{#1}}}
\def\author#1{\gdef\@author{\uppercase{#1}}}
\def\abstract#1{\gdef\@abstract{#1}}
\def\journal#1{\gdef\@journal{Submitted to {\em #1}}}
\def\address#1{\gdef\@address{{\em #1}}}

\def\makefrontmatter{
	\thispagestyle{empty}\pagenumbering{roman}%
	\begin{center}
	\rightline{astro-ph/9610224}\vskip2ex
	\null\vfill
	\@title
	\vfill
	\@author
	\vskip 2ex
	\@address
	\end{center}
	\vfill
	\centerline{\bf Abstract}\vskip 2ex
	\@abstract
	\vfill
	\noindent\@journal
	\newpage
	\pagenumbering{arabic}
	\setcounter{page}{1}
	\thispagestyle{plain}
}

\def\section{\@startsection{section}{1}{\z@}{-1.5ex}{1.5ex}%
{\large\bf}}
\def\subsection{\@startsection{subsection}{2}{\z@}{-1.2ex}{1.2ex}%
{\normalsize\bf}}
\def\subsubsection{\@startsection{subsubsection}{3}{\z@}{-1ex}{1ex}%
{\small\bf}}

\def\references{\clearpage\thispagestyle{plain}\section*{References}
	\addcontentsline{toc}{section}{References}
	\bgroup\parindent=\z@\parskip=\itemsep
	\frenchspacing 
	\def\ibid{\vrule height3pt depth-2.6pt width3em}
	\def\refpar{\par\hangindent=3em\hangafter=1}}
\def\endreferences{\refpar\egroup}
\def\reference{\relax\refpar}

\makeatother

  \evensidemargin=\oddsidemargin
\title{A Stochastic Approach to Chemical Evolution}
\author{Craig J. Copi}
\address{Department of Physics\\ Enrico Fermi Institute, The University of
Chicago, Chicago, IL~~60637-1433 \\*[1em]
NASA/Fermilab Astrophysics Center \\ Fermi National Accelerator Laboratory,
Batavia, IL~~60510-0500}
\journal{The Astrophysical Journal}

\newif\ifincludeeps
\includeepstrue
\newdimen\figwidth \figwidth=6in

\makeatletter
\def\deffig#1#2{\@namedef{fig@#1}{\begin{figure} #2 \end{figure}}}
\def\putfig#1{\@nameuse{fig@#1}}
\makeatother

\deffig{age-metallicity}{
\ifincludeeps \epsfysize=\figwidth\center\leavevmode
\rotate[r]{\epsfbox{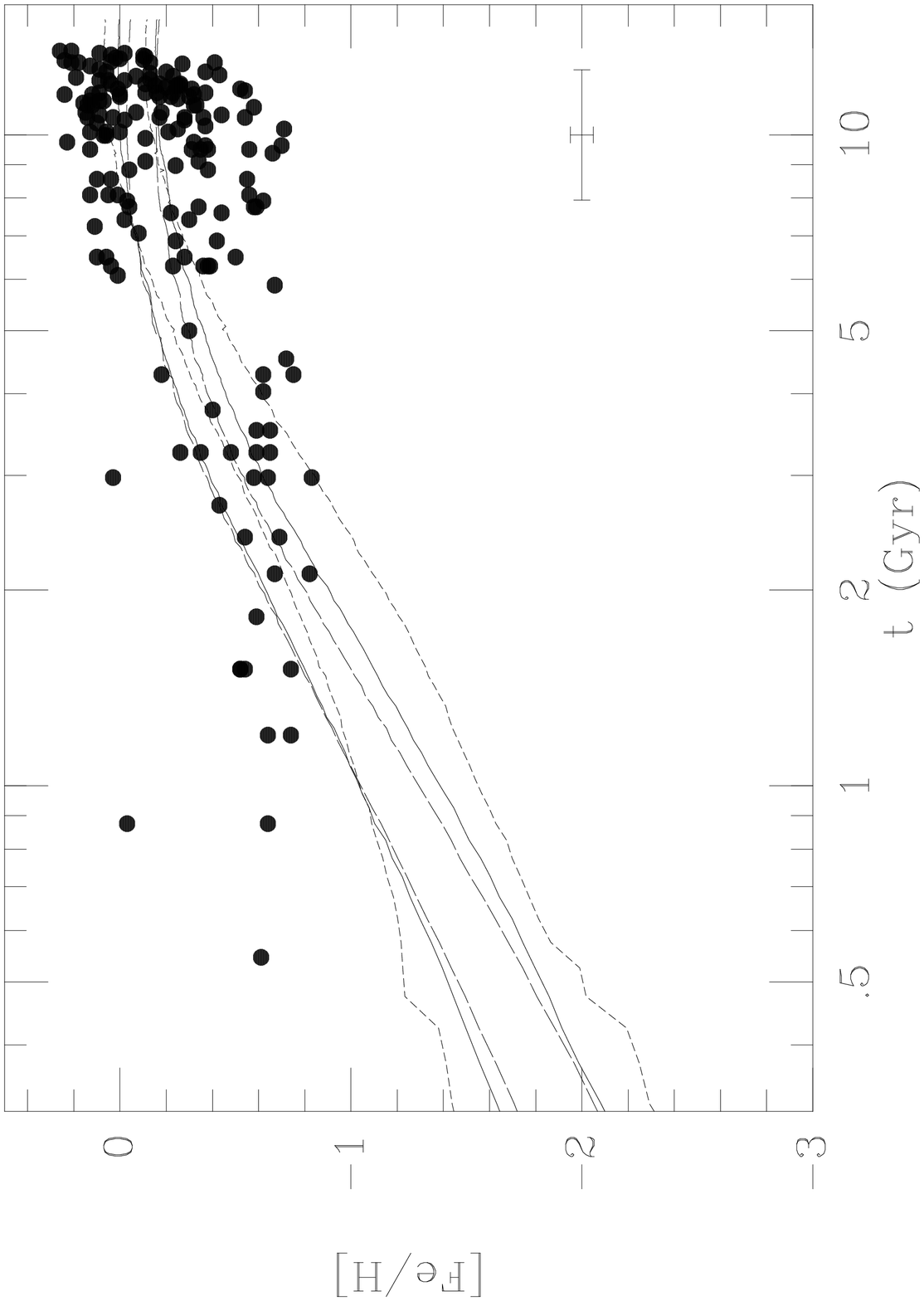}}
\fi
\caption[Age-metallicity relation]{The age-metallicity relation for the three
models considered in this work: the closed box (solid line), infall
(short-dashed line), and outflow (long-dashed line) models.
The data is from Edvardsson \etal~(1993).  A typical error bar is shown in the
bottom right corner of the figure.  Note that none of the models accurately
reproduce the spread in the observations.}
\label{fig:age-metallicity}
}

\deffig{heavy-solar}{
\ifincludeeps \epsfysize=\figwidth\center\leavevmode
\rotate[r]{\epsfbox{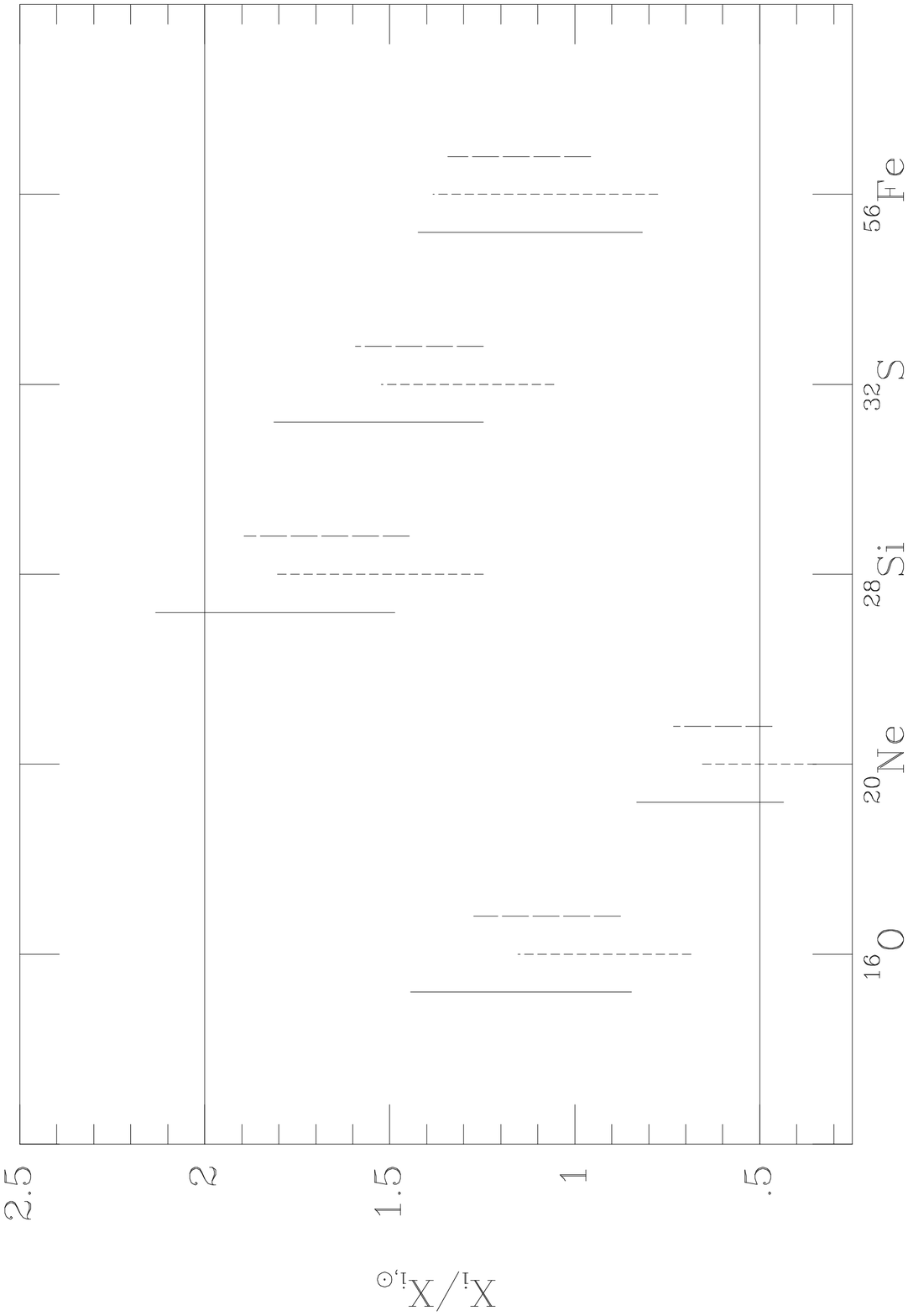}}
\fi
\caption[Heavy element solar abundance ratios]{The solar abundance ratios for
the heavy elements.  The ranges within which 95\% of the models fall for
the closed box (solid line), infall (short-dashed line), and outflow
(long-dashed line) models.
The solar data comes from Anders \& Grevesse~(1989).  Note that \Fe{56} was
used to fix the type Ia supernova rate so its good agreement is required.  We
allow ourselves a factor of two uncertainty when comparing the predictions and
observations as discussed in the text.}
\label{fig:heavy-solar}
}

\deffig{O-Fe}{
\ifincludeeps \epsfysize=\figwidth\center\leavevmode
\rotate[r]{\epsfbox{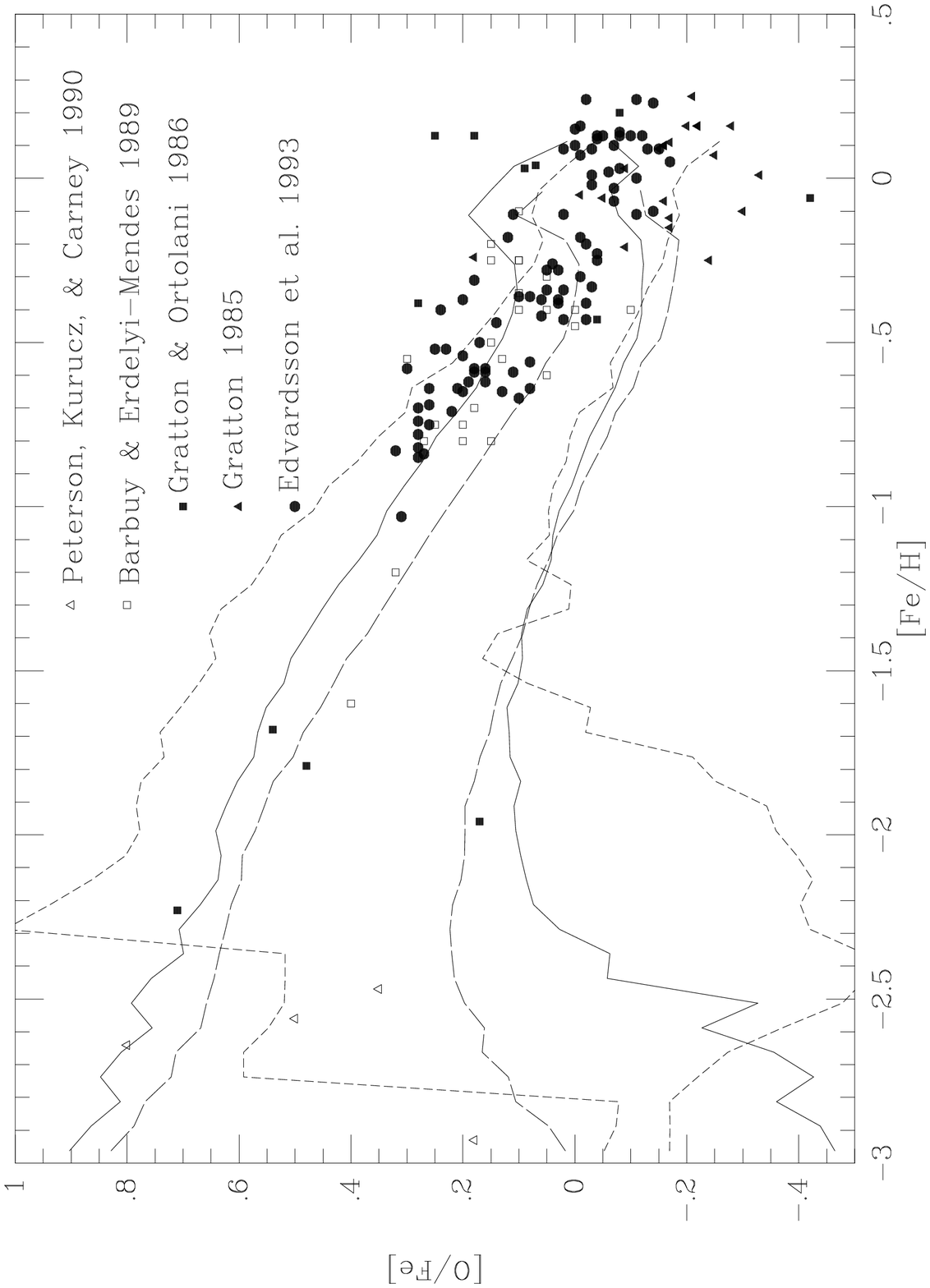}}
\fi
\caption[The oxygen-to-iron ratio as a function of iron abundance]{The
oxygen-to-iron ratio as a function of iron abundance for the three models
considered.  Since oxygen is not sensitive to the initial abundance nor the
low mass stellar yields employed, the results shown for the closed box (solid
line), infall (short-dashed line), and outflow (long-dashed line) models are
generic to the model. Notice that the predicted spread is in good agreement
with observation, particularly for iron abundances, $[\feh] \ltsim -1$.}
\label{fig:O-Fe}
}

\deffig{Ne-Fe}{
\ifincludeeps \epsfysize=\figwidth\center\leavevmode
\rotate[r]{\epsfbox{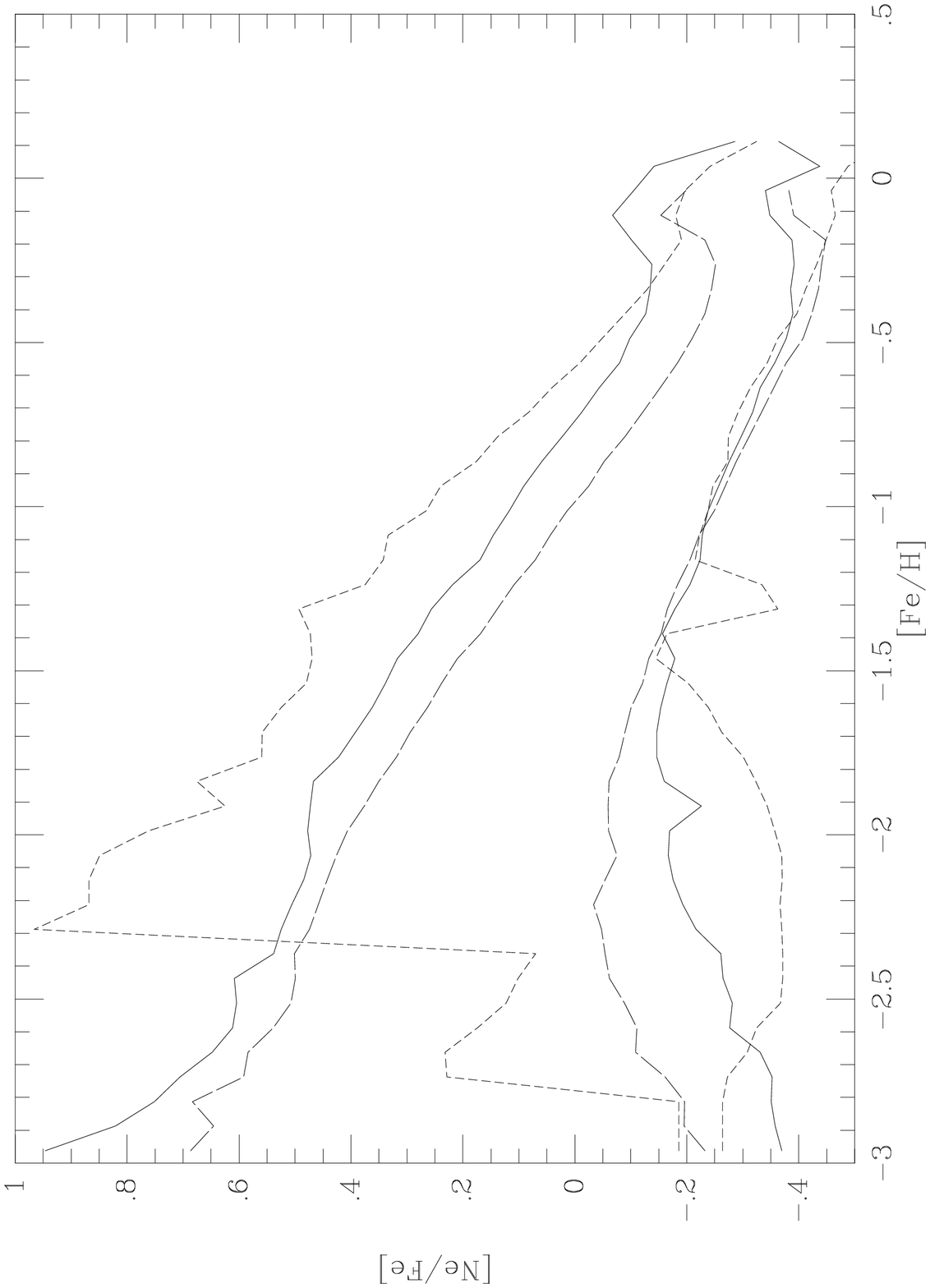}}
\fi
\caption[The neon-to-iron ratio as a function of iron abundance]{The
neon-to-iron ratio as a function of iron abundance for the three models
considered.  See figure~\protect\ref{fig:O-Fe} for details.}
\label{fig:Ne-Fe}
}

\deffig{Si-Fe}{
\ifincludeeps \epsfysize=\figwidth\center\leavevmode
\rotate[r]{\epsfbox{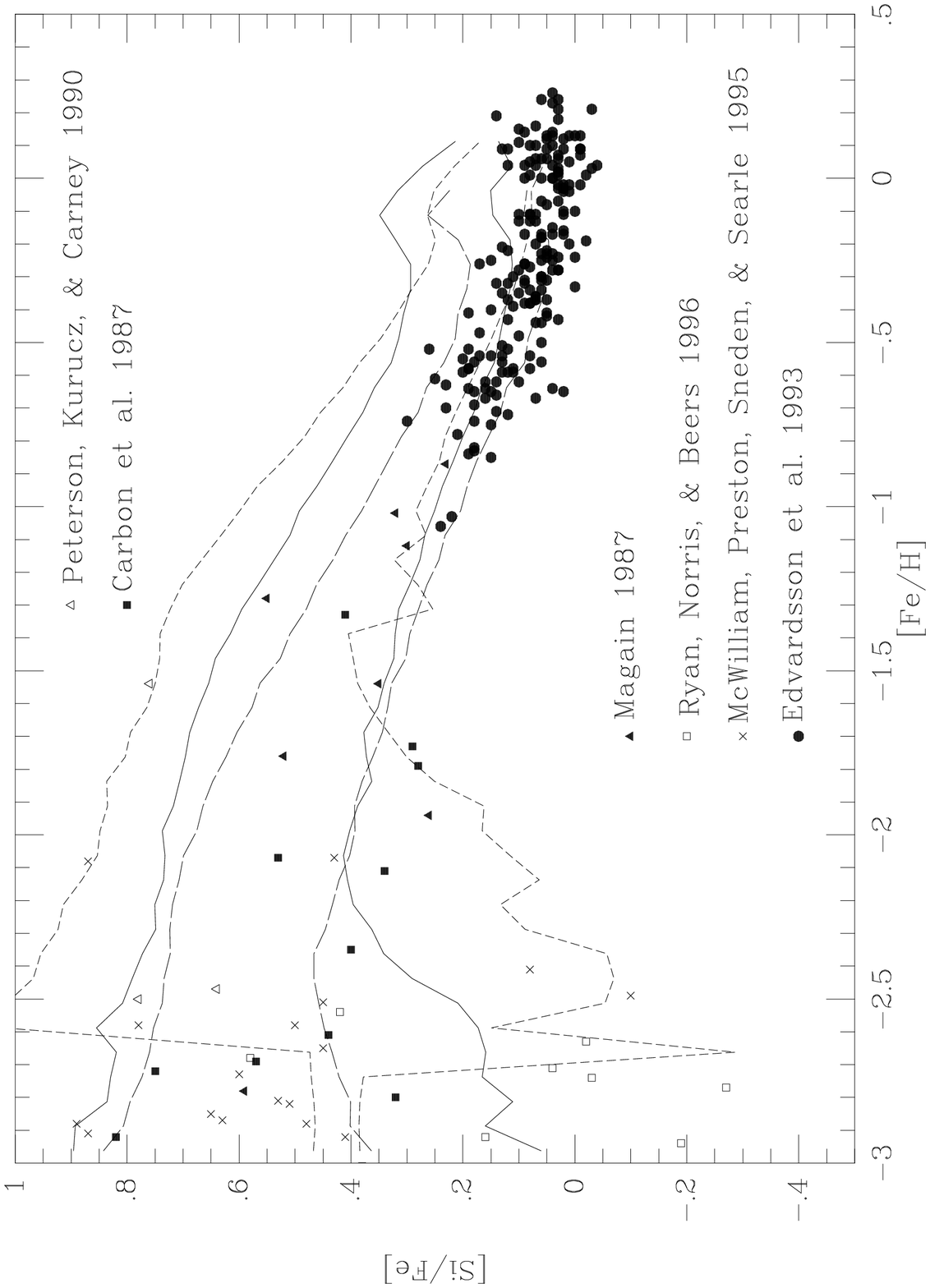}}
\fi
\caption[The silicon-to-iron ratio as a function of iron abundance]{The
silicon-to-iron ratio as a function of iron abundance for the three models
considered.  See figure~\protect\ref{fig:O-Fe} for details.}
\label{fig:Si-Fe}
}

\deffig{S-Fe}{
\ifincludeeps \epsfysize=\figwidth\center\leavevmode
\rotate[r]{\epsfbox{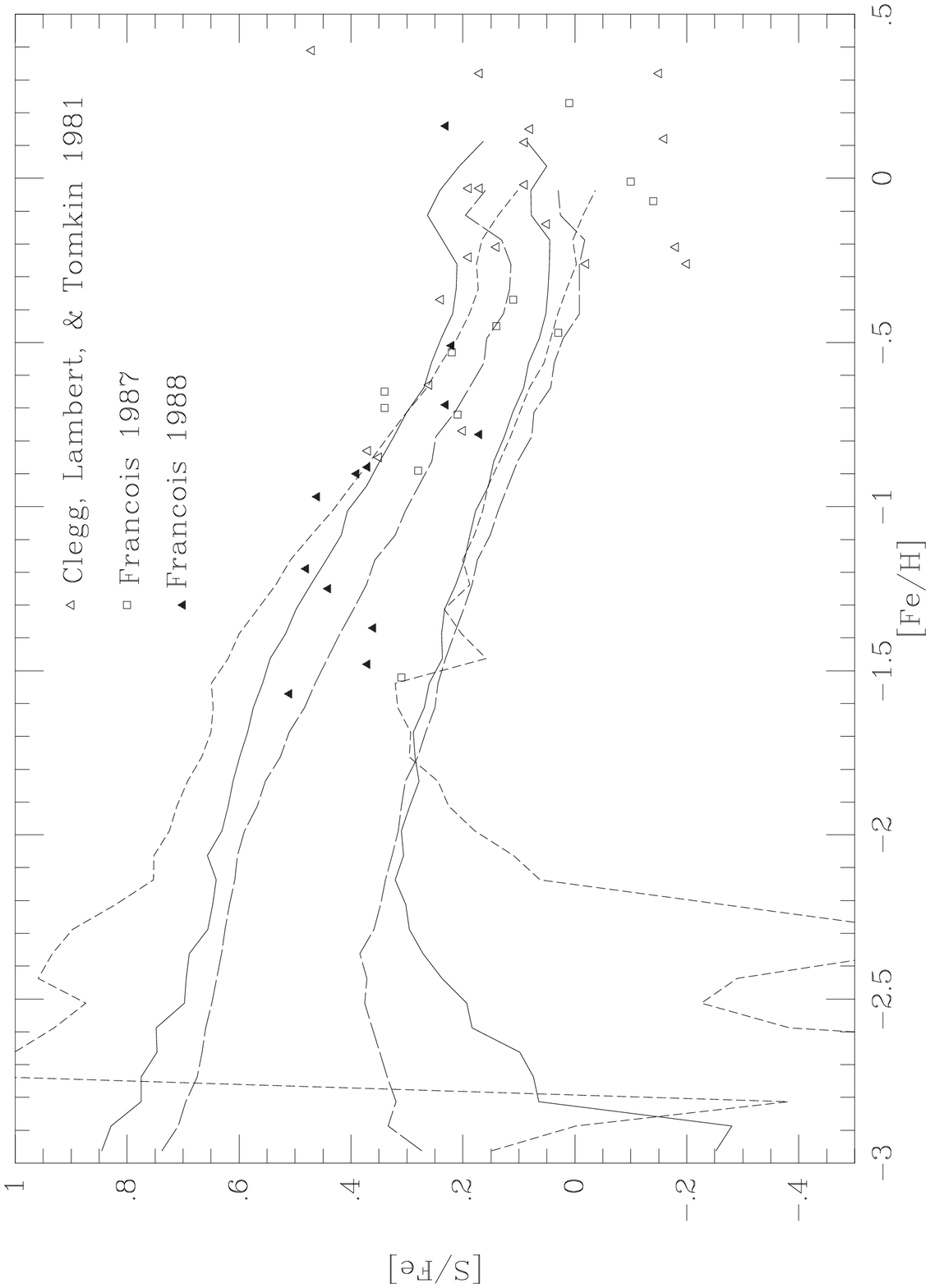}}
\fi
\caption[The sulfur-to-iron ratio as a function of iron abundance]{The
sulfur-to-iron ratio as a function of iron abundance for the three models
considered.  See figure~\protect\ref{fig:O-Fe} for details.}
\label{fig:S-Fe}
}

\deffig{CN-solar}{
\ifincludeeps \epsfysize=\figwidth\center\leavevmode
\rotate[r]{\epsfbox{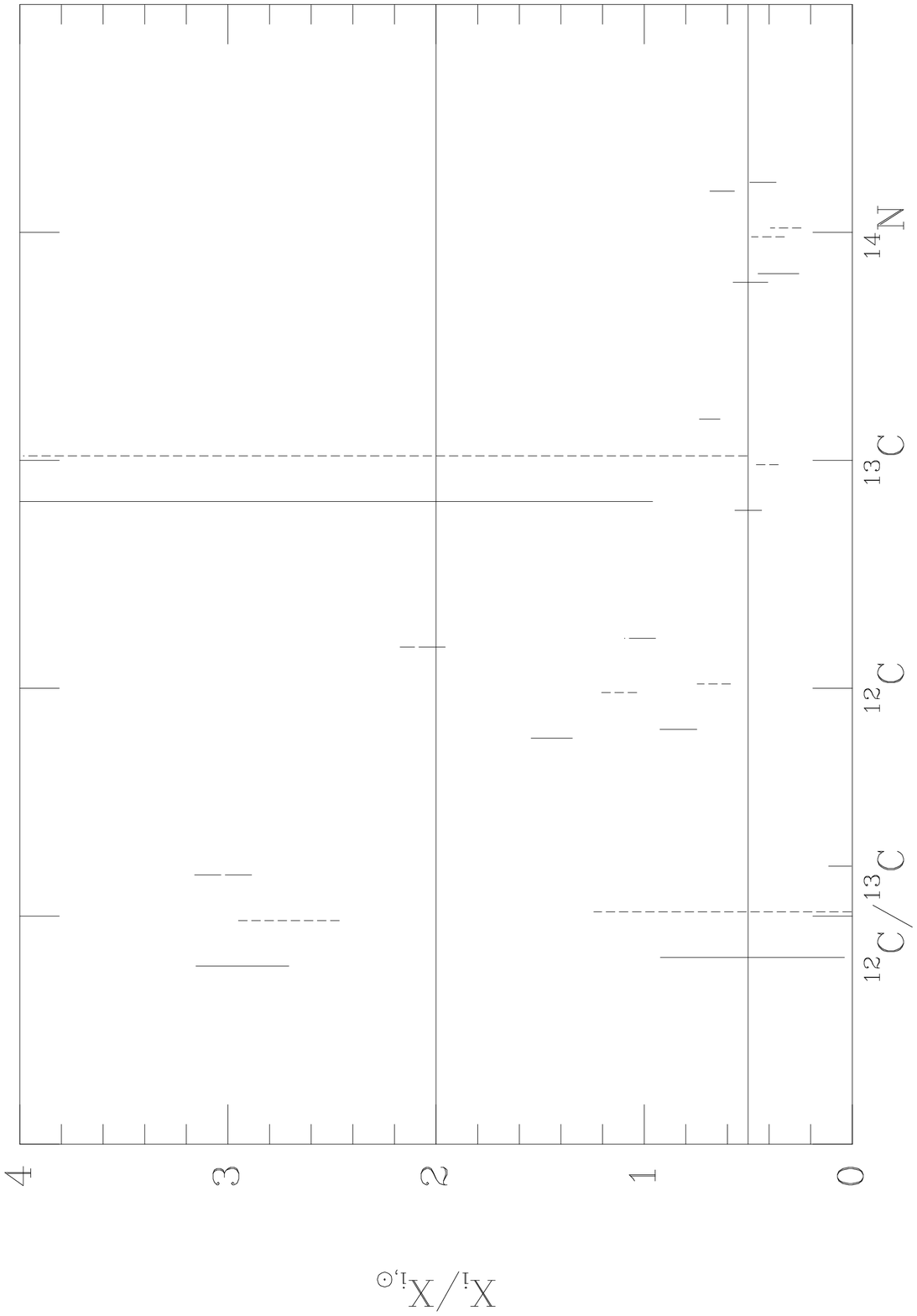}}
\fi
\caption[Carbon and nitrogen solar ratio abundances]{The solar abundance
ratios for the carbon and nitrogen isotopes.  The ranges are as given in
figure~\protect\ref{fig:heavy-solar}.  Since the carbon and nitrogen
abundances depend on the low mass yields chosen we show two sets of results
for each model depending on our choice of low mass yields.  The
yields with extra mixing are the lower set of ranges for \C{12} and the
\C{12}/\C{13} ratio, the upper set of ranges for \C{13}, and the ranges offset
slightly to the right for \N{14}.  See the text for details.}
\label{fig:CN-solar}
}

\deffig{C-Fe}{
\ifincludeeps \epsfxsize=0.9\figwidth \center\leavevmode
{\epsfbox{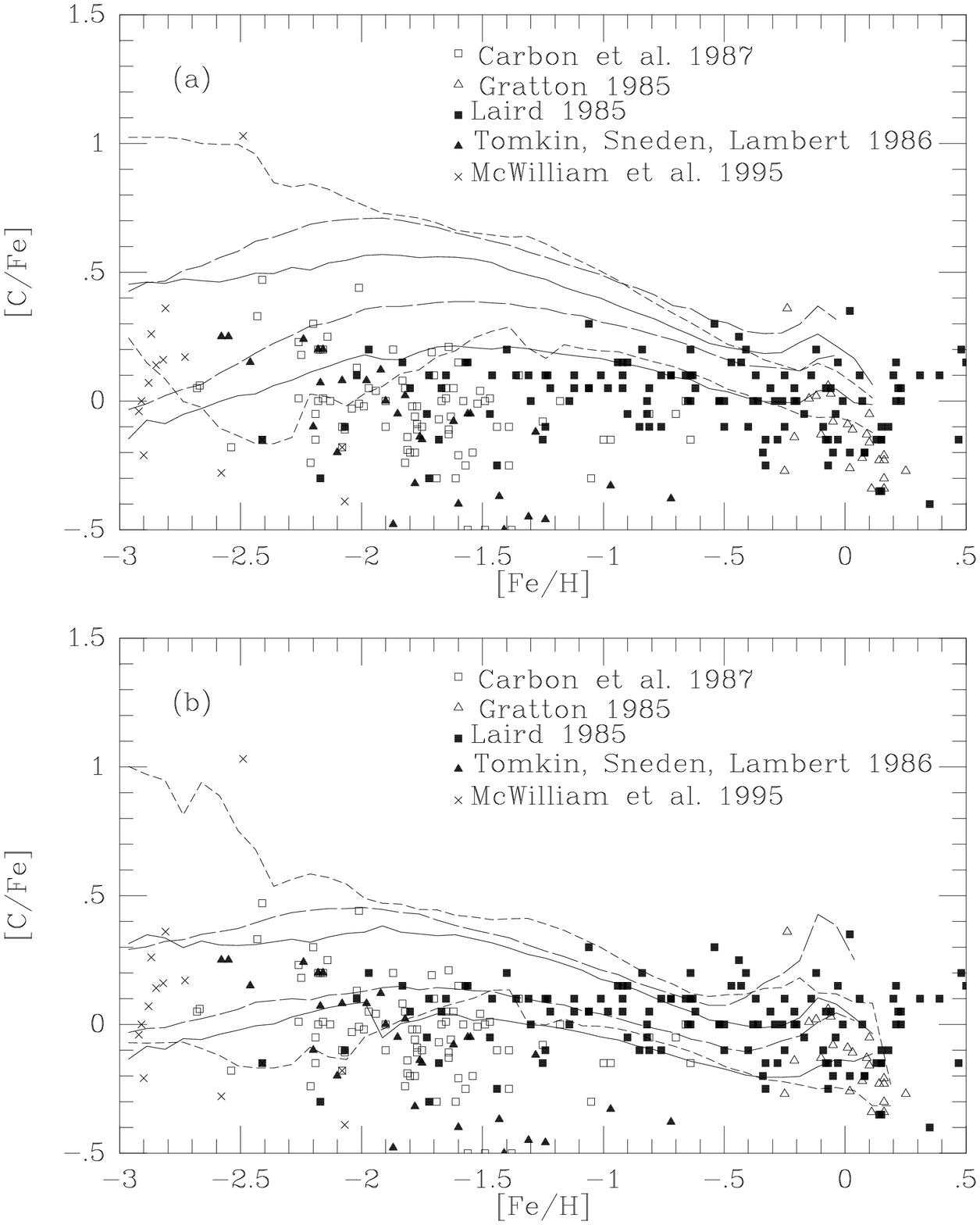}}
\fi
\caption[The carbon-to-iron ratio as a function of the iron abundance]{The
carbon-to-iron abundance as a function of iron abundance for the three models
considered.  The labels are as in figure~\protect\ref{fig:O-Fe}.  In (a) we
show the results from the older low mass stellar yields and in (b) we show the
results from the newer yields that include extra mixing.  Notice that only
about a half of the scatter in the data can be explained by the different
histories of regions.  See the text for more details.}
\label{fig:C-Fe}
}

\deffig{N-Fe}{
\ifincludeeps \epsfysize=\figwidth\center\leavevmode
\rotate[r]{\epsfbox{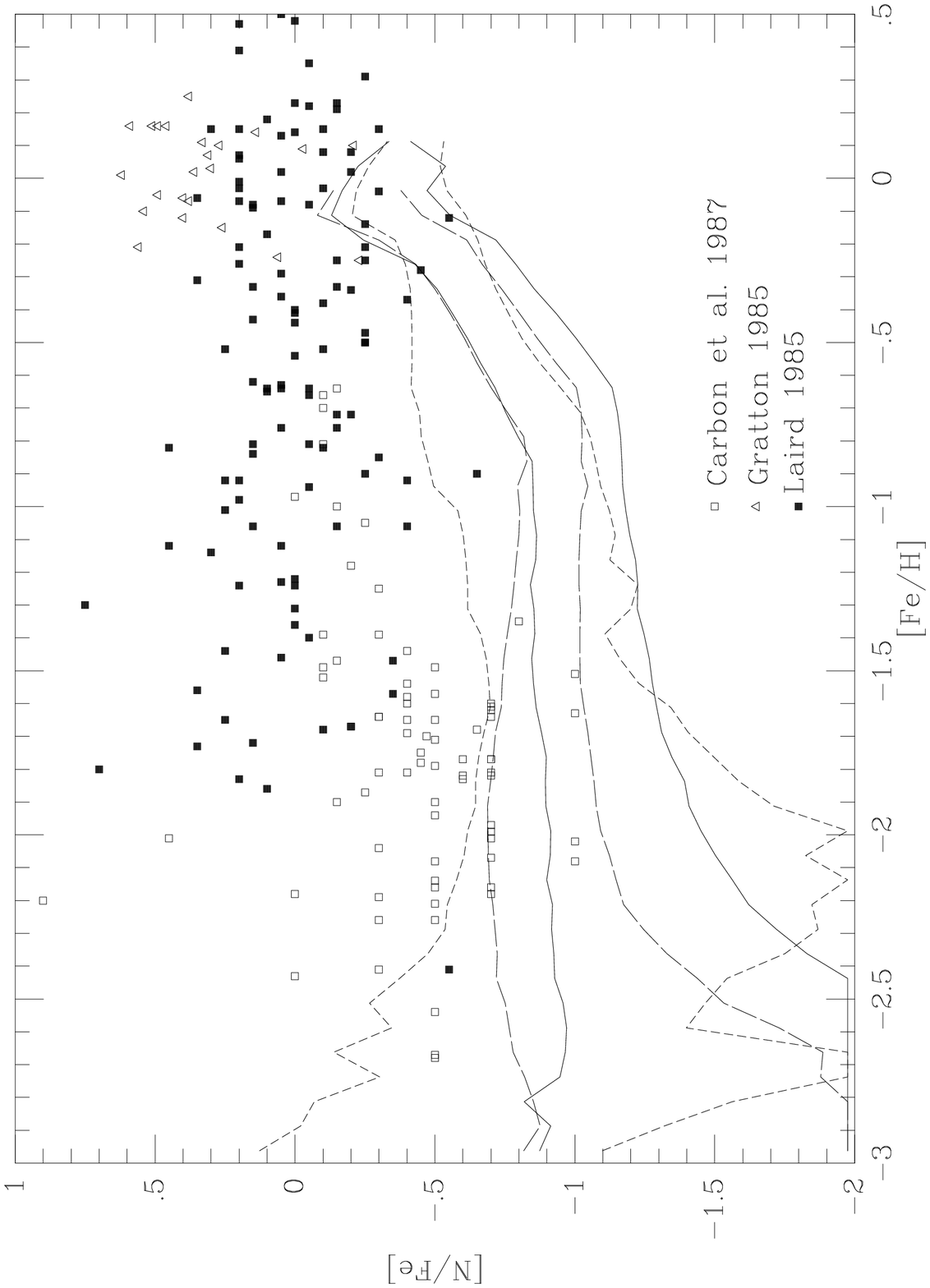}}
\fi
\caption[The nitrogen-to-iron ratio as a function of the iron abundance]{The
nitrogen-to-iron abundance as a function of iron abundance for the three models
considered.  The labels are as in figure~\protect\ref{fig:O-Fe}.  We only show
the results for the older yields since the results with the newer yields are
quite similar.  Similar to the case with carbon
(figure~\protect\ref{fig:C-Fe}) we see a large scatter in the data.  See the
text for details.} 
\label{fig:N-Fe}
}

\deffig{lihe-solar}{
\ifincludeeps \epsfysize=\figwidth\center\leavevmode
\rotate[r]{\epsfbox{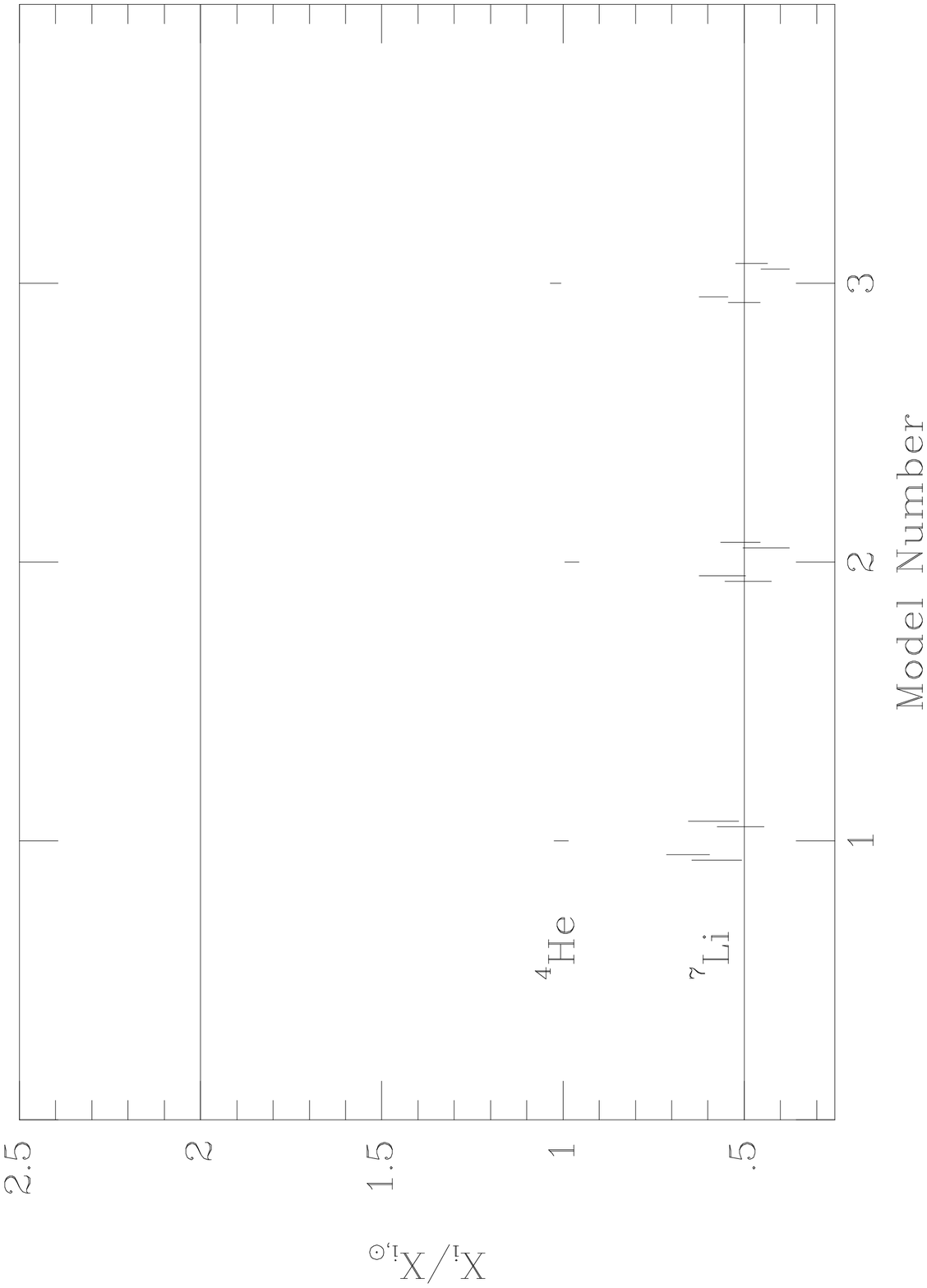}}
\fi
\caption[The solar abundance ratios for \protect\Li{7} and \protect\He{4}]
{The solar abundance ratios for \protect\Li{7} and \protect\He{4}.  The
lithium abundance is dependent on both the initial abundance and on the low
mass stellar yields,
thus four ranges are shown for each model.  The two lower ranges are for $\eta
= 4.5\sci{-10}$ and the upper ranges are for $\eta =
5.5\sci{-10}$.  Similarly the left two ranges are for the older
low mass stellar yields and the right two ranges are for the
newer yields.  In principle the \He4\ abundance also depends on
our choice of initial abundance and low mass stellar yields.  In practice this
dependence is found to be extremely small so only one range is shown for each
model.  See the text for details.}
\label{fig:lihe-solar}
}

\deffig{Li}{
\ifincludeeps \epsfysize=\figwidth\center\leavevmode
\rotate[r]{\epsfbox{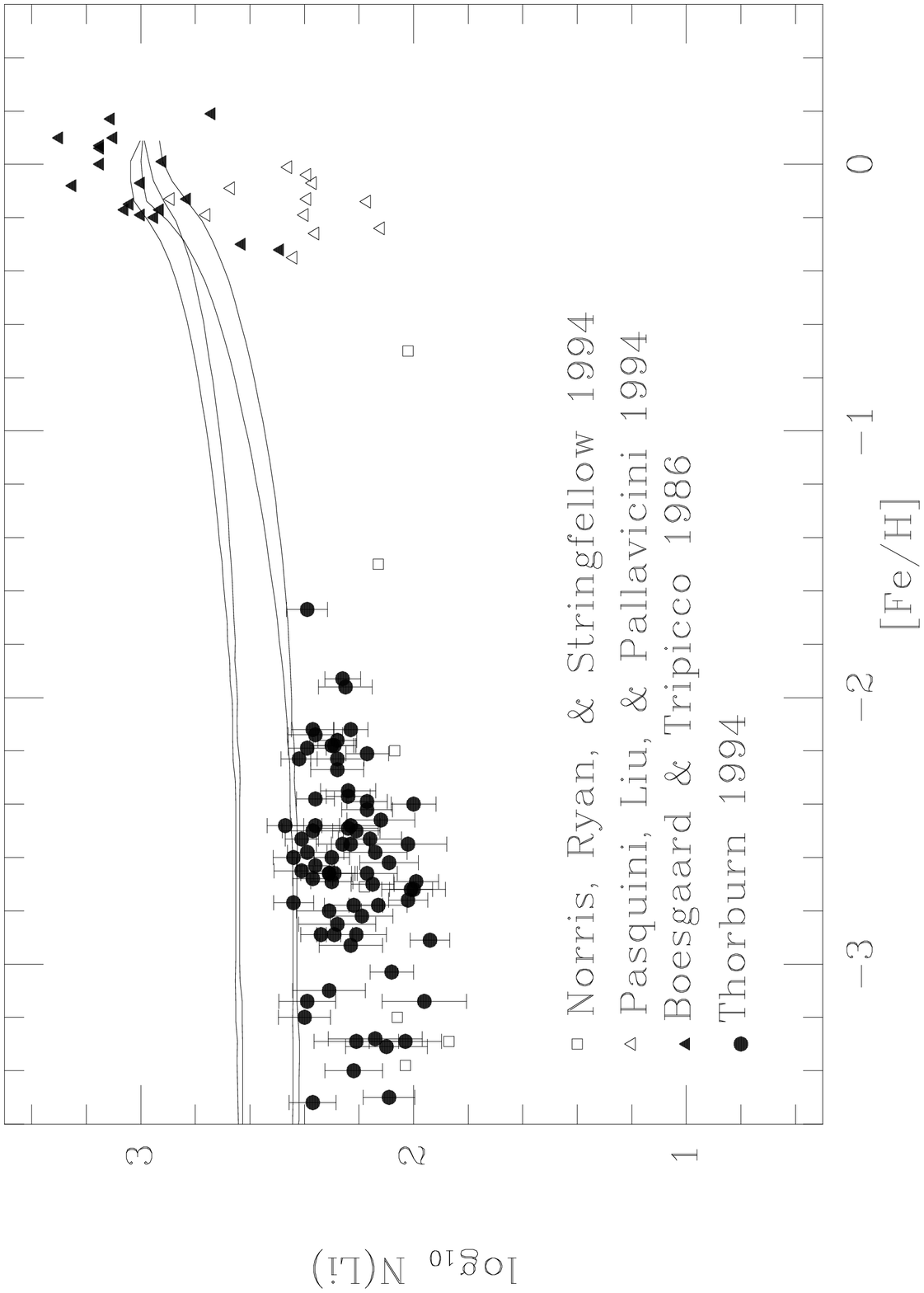}}
\fi
\caption[The lithium abundance as a function of the iron abundance]{The
lithium as a function of iron abundance for the closed box model.  All three
models are quite similar, particularly on the Spite plateau.  We only show
the results for the older yields since the results with the newer
yields are quite similar with only a slight vertical shift.  The lower set of
curves are for $\eta=4.5\protect\sci{-10}$ and the upper set of curves are for
$\eta=5.5\protect\sci{-10}$. See the text for more details.}
\label{fig:Li}
}

\deffig{Y-O}{
\ifincludeeps \epsfysize=\figwidth\center\leavevmode
\rotate[r]{\epsfbox{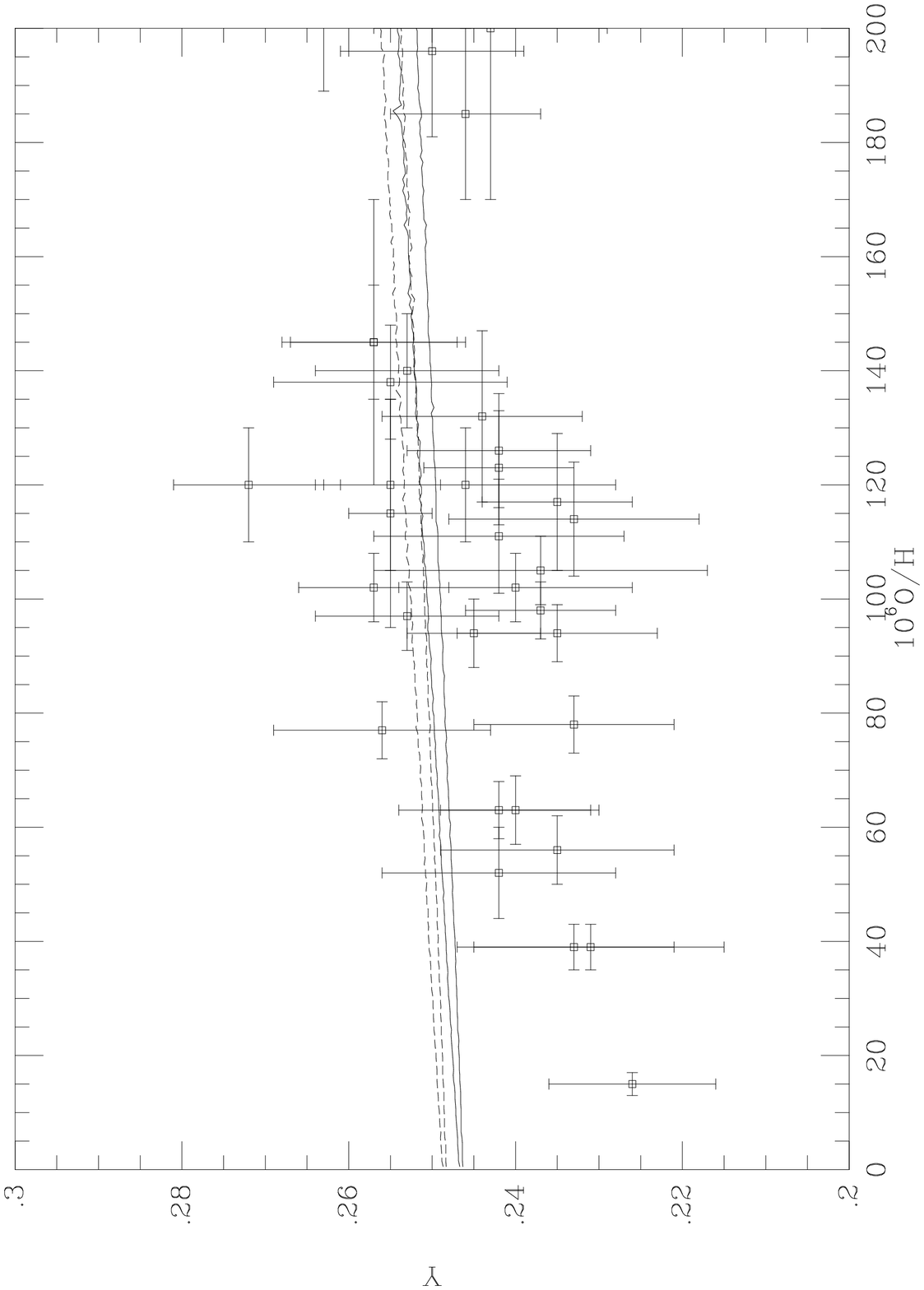}}
\fi
\caption[The \protect\He4\ abundance as a function of the oxygen abundance]{The
\protect\He4\ abundance as a function of the oxygen abundance.  The data is
from Pagel \etal~(1992).  Shown here are the results for the closed box model
with $\eta=4.5\protect\sci{-10}$ (solid
line) and $\eta=5.5\protect\sci{-10}$ (short-dashed line). Results for the
other models are nearly identical.  See the text for details.}
\label{fig:Y-O}
}

\deffig{D-He-v-t}{
\ifincludeeps \epsfysize=0.5\figwidth\center\leavevmode
\hbox{\rotate[r]{\epsfbox{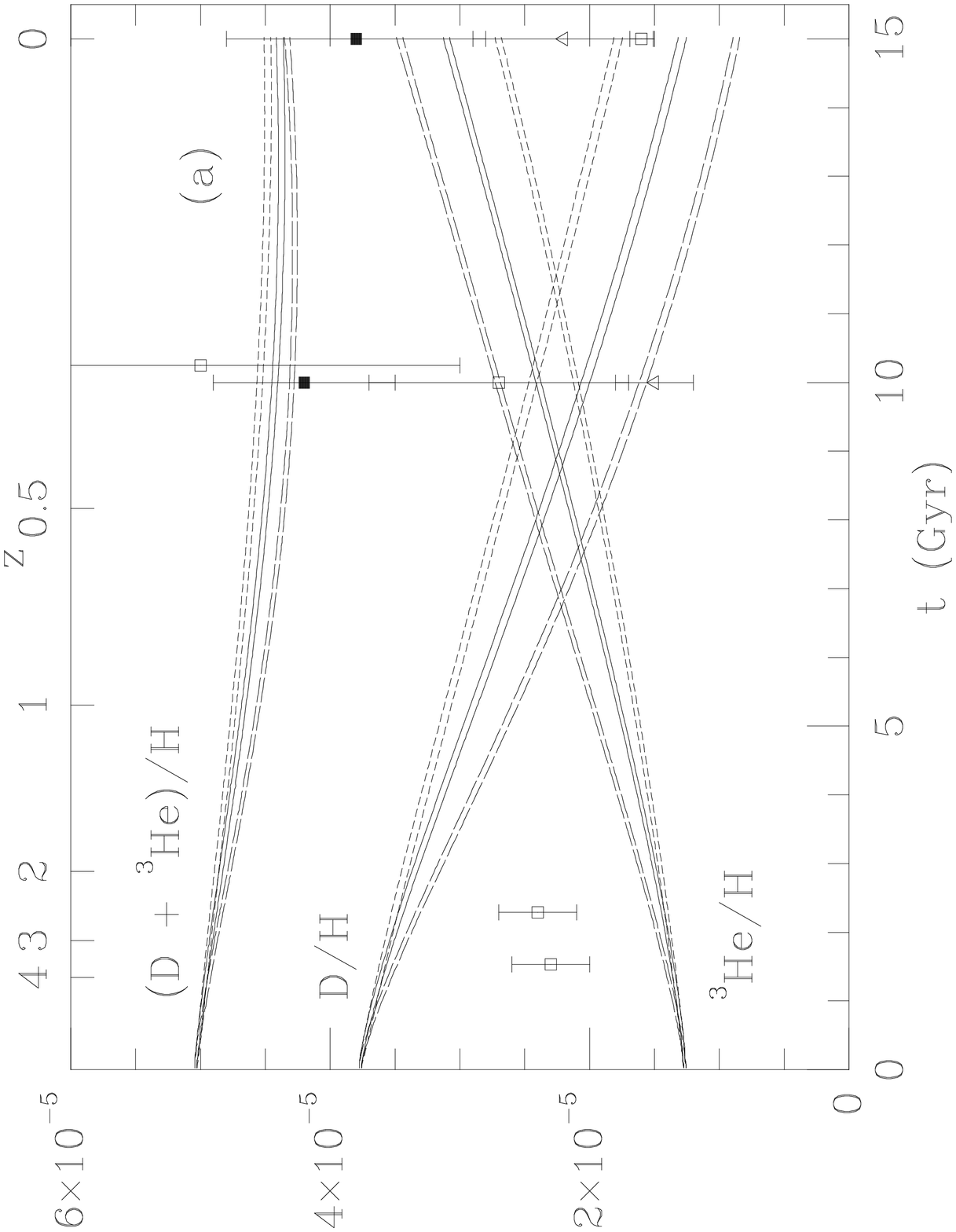}}\hskip1em\rotate[r]{\epsfbox{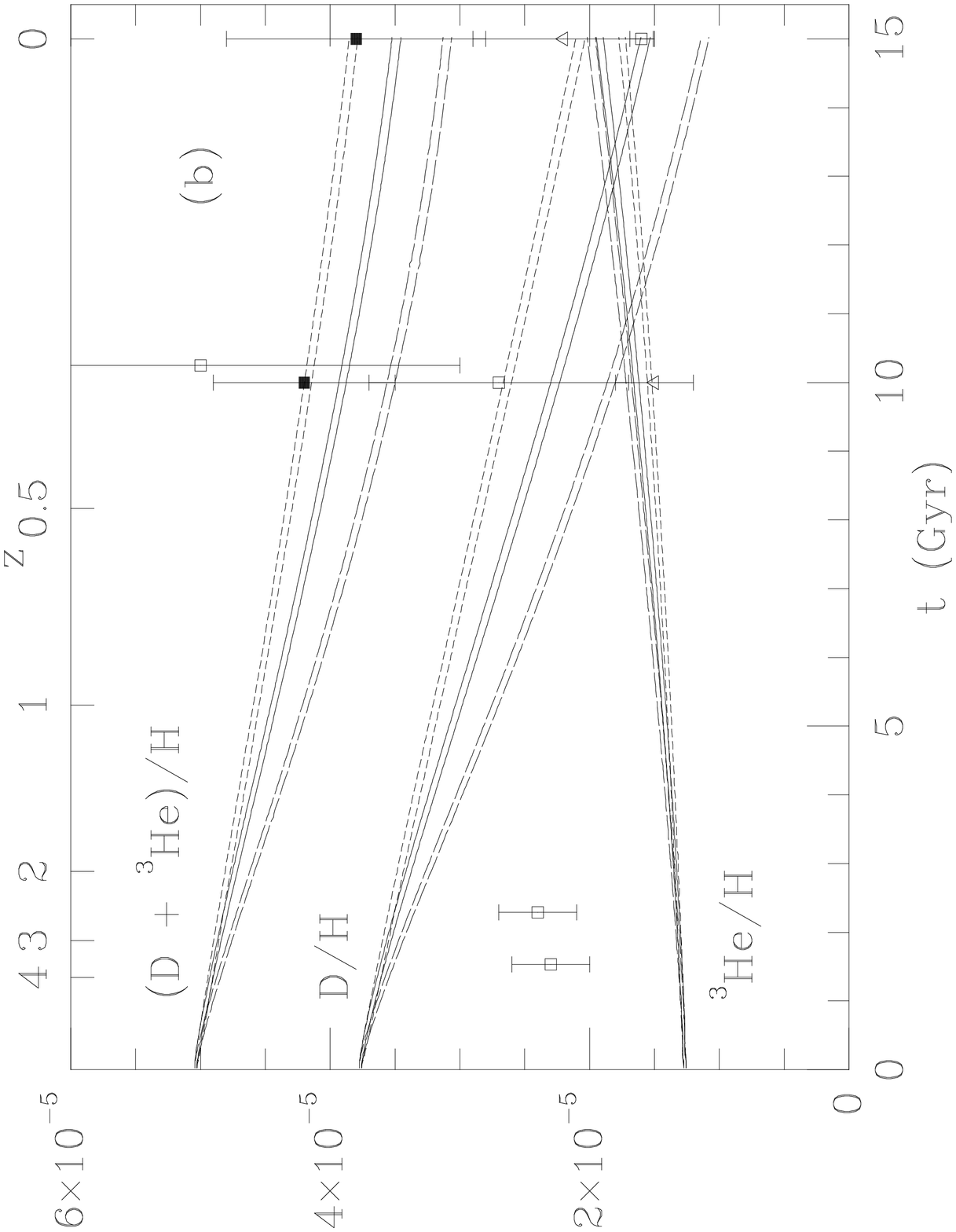}}}
\vskip 2ex\leavevmode
\hbox{\rotate[r]{\epsfbox{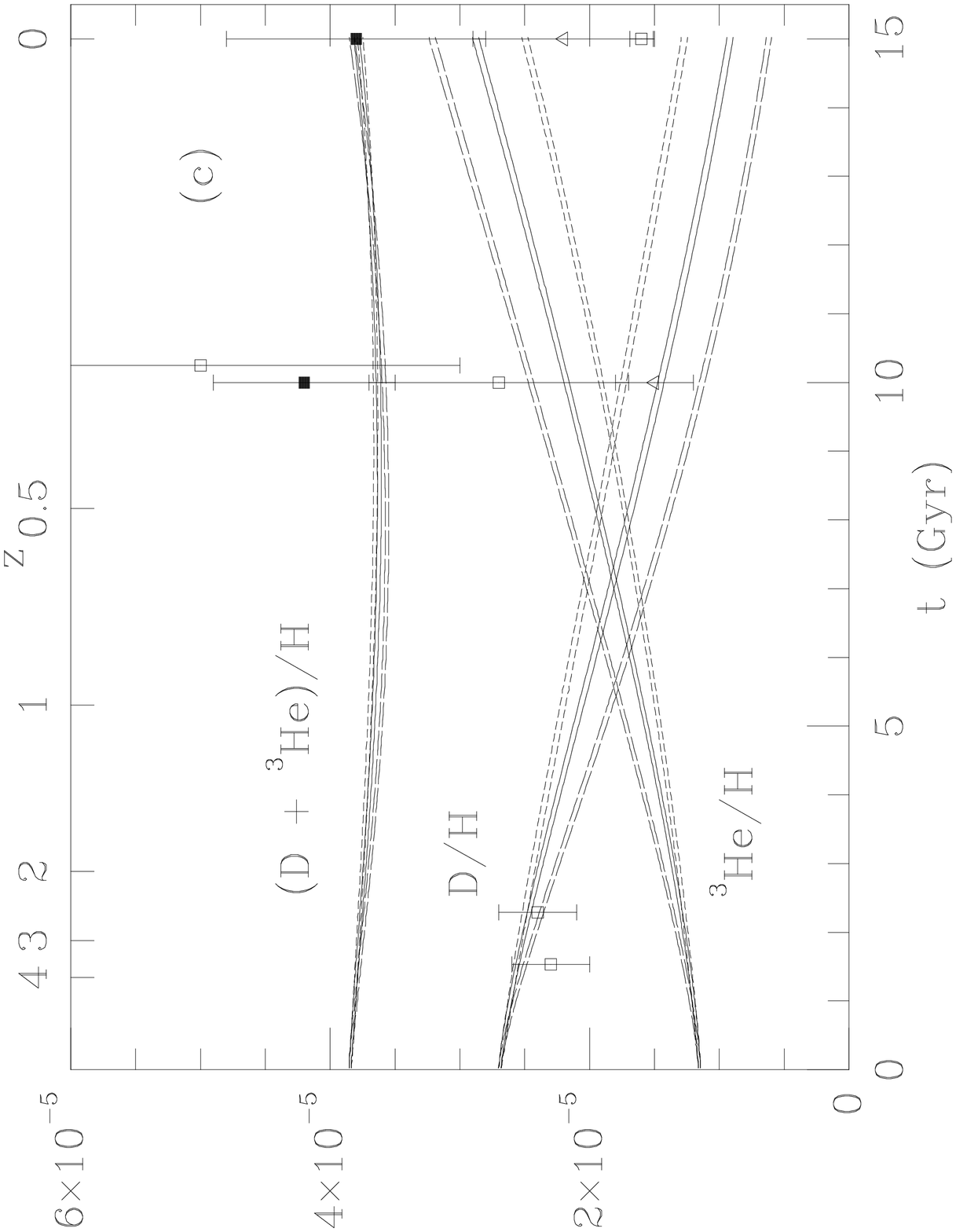}}\hskip1em\rotate[r]{\epsfbox{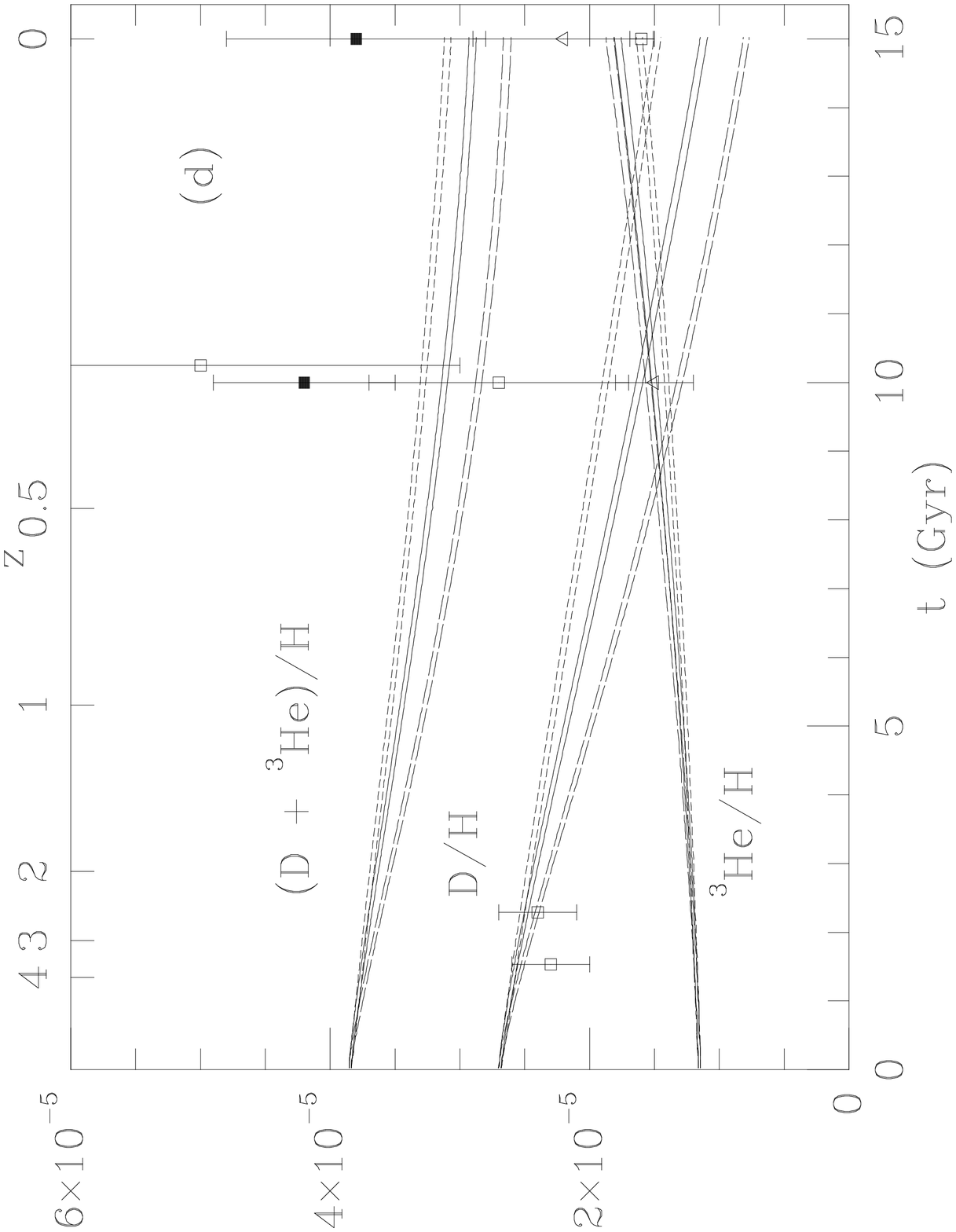}}}
\fi
\caption[Time evolution of D and \protect\He3]{The time evolution of D/H,
\protect\He3/H, and (D+\protect\He3)/H for the 3 models.  The line types are
as in figure~\protect\ref{fig:O-Fe}.  The data for D
($\Box$), \protect\He{3} ($\triangle$), and  D+\protect\He{3}
($\solidBox$) are shown.  The observation of D in the atmosphere of Jupiter
(Niemann \etal~1996) is shown shifted slightly to the right of the pre-solar
observations for clarity.  The redshift scale is for a flat, $\Omega_{\rm
Matter}=1$, Universe with an age of 15\protect\Gyr.  The panels represent the
results for (a)~models with $\eta=4.5\protect\sci{-10}$ and \protect\He{3}
production from Iben \& Truran~(1978), (b)~same as (a) except we use the newer
low mass stellar yields from Boothroyd \& Sackmann~(1996), (c)~same as (a) for
$\eta=5.5\protect\sci{-10}$, and (d)~same as (b) for
$\eta=5.5\protect\sci{-10}$.}
\label{fig:D-He-v-t}
}

\deffig{D-ja}{
\ifincludeeps \epsfysize=\figwidth\center\leavevmode
\rotate[r]{\epsfbox{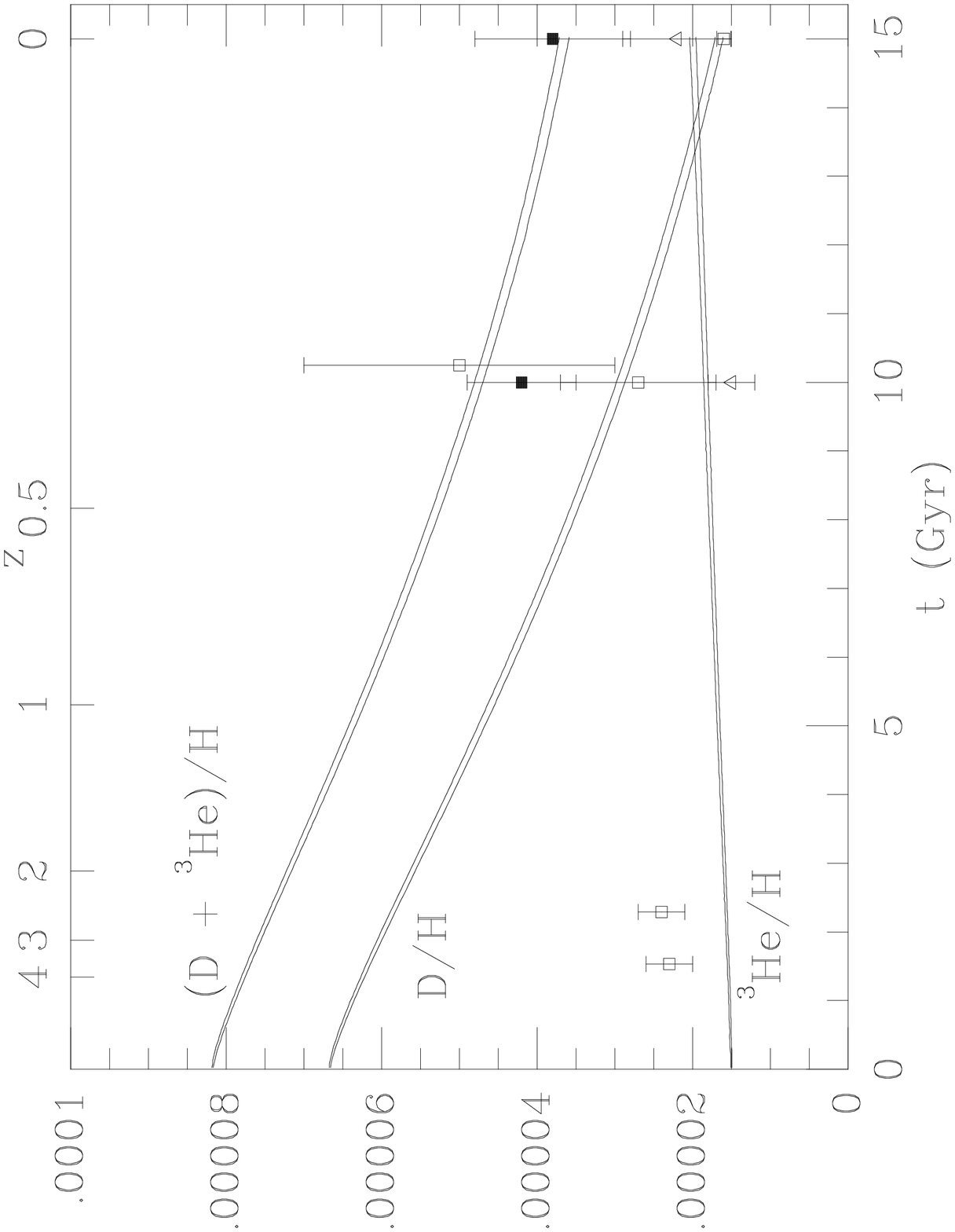}}
\fi
\caption[Time evolution of D and \protect\He{3} for
$\eta=3.2\protect\sci{-10}$]{The time evolution of D/H,
\protect\He3/H, and (D+\protect\He3)/H for an outflow model with
$\eta=3.2\protect\sci{-10}$.  See the text for details.  The data is as in
figure~\protect\ref{fig:D-He-v-t}.}
\label{fig:D-ja}
}

\deffig{pdmf}{
\ifincludeeps \epsfysize=\figwidth\center\leavevmode
\rotate[r]{\epsfbox{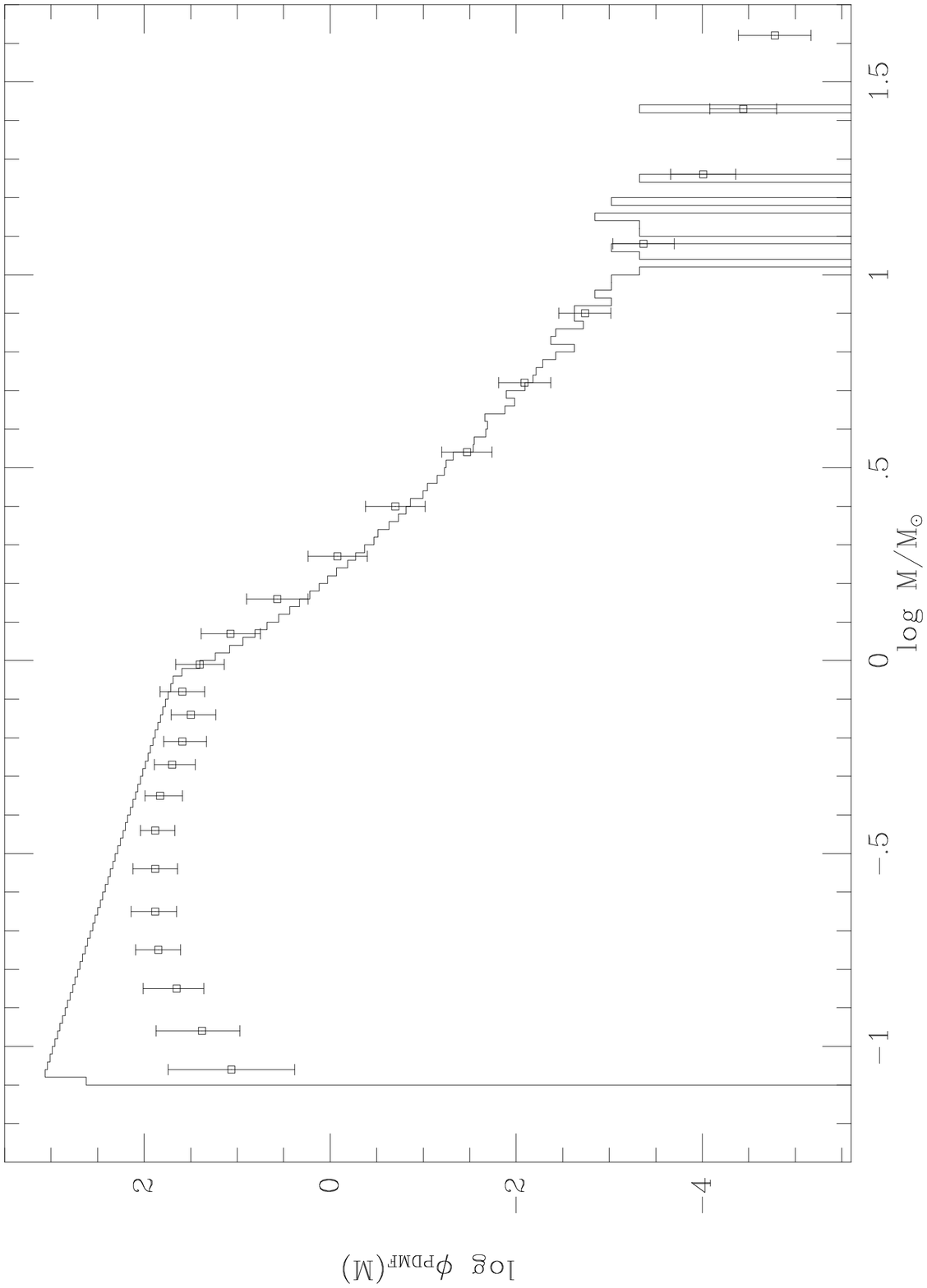}}
\fi
\caption[Present day mass function]{The distribution of stars, by mass,
expected to currently reside in the solar neighborhood, known as the present
day mass function.  Only the curve for the closed box model
is shown since all models are nearly identical.  The data is from
Scalo~(1986).  Since we have employed a power law IMF we do not expect to get
good agreement at low masses, $\log M \ltsim 0$.}
\label{fig:pdmf}
}

\deffig{gdwarf}{
\ifincludeeps \epsfysize=\figwidth\center\leavevmode
\rotate[r]{\epsfbox{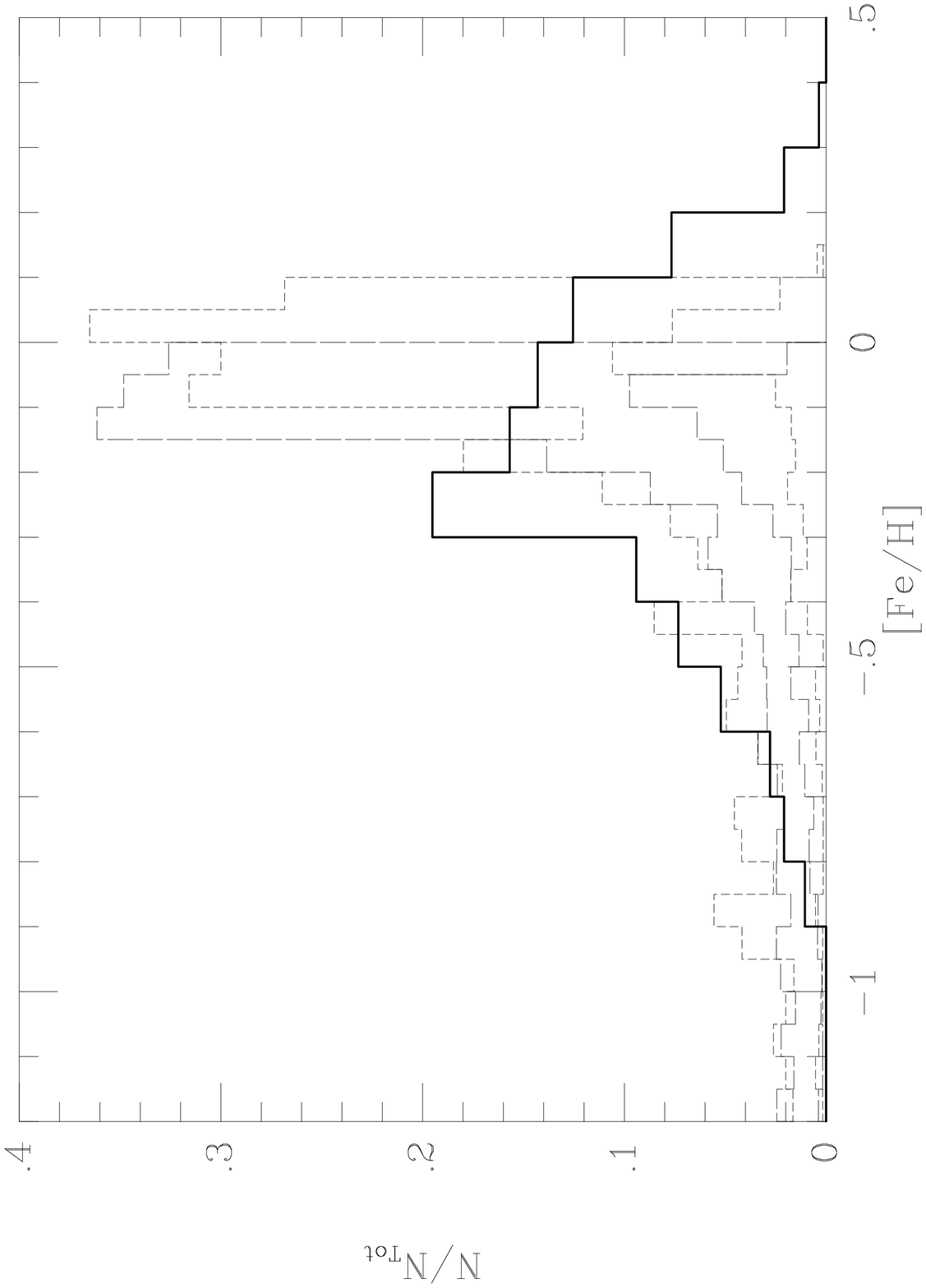}}
\fi
\caption[G-dwarf distribution]{The spread in the number of G-dwarf stars as a
function of iron abundance.  The data (heavy solid line) is from Rocha-Pinto \&
Maciel~(1996).  The line types are as in figure~\protect\ref{fig:O-Fe}.  The
results for the closed box model are left out for clarity.  The results for
the closed box model falls between the infall and outflow models.
}
\label{fig:gdwarf}
}

\abstract{ Observations of elemental abundances in the Galaxy have repeatedly
shown an intrinsic scatter as a function of time and metallicity.  The
standard approach to chemical evolution does not attempt to address this
scatter in abundances since only the mean evolution is followed.  In this work
the scatter is addressed via a stochastic approach to solving chemical
evolution models.  Three standard chemical evolution scenarios are studied
using this stochastic approach; a closed box model, an infall model, and an
outflow model.  These models are solved for the solar neighborhood in a Monte
Carlo fashion.  The evolutionary history of one particular region is
determined randomly based on the star formation rate and the initial mass
function.  Following the evolution in an ensemble of such regions leads to the
predicted spread in abundances expected, based solely on different
evolutionary histories of otherwise identical regions.  In this work 13
isotopes are followed including the light elements, the CNO elements, a few
$\alpha$-elements, and iron.  It is found that the predicted spread in
abundances for a $10^5\Msun$ region is in good agreement with observations for
the $\alpha$-elements.  For CN the agreement is not as good perhaps indicating
the need for more physics input for low mass stellar evolution.  Similarly for
the light elements the predicted scatter is quite small which is in
contradiction to the observations of \He{3} in \hii\ regions.  The models are
tuned for the solar neighborhood so good agreement with \hii\ regions is not
expected.  This has important implications for low mass stellar evolution and
on using chemical evolution to determine the primordial light element
abundances in order to test big-bang nucleosynthesis.  }

\begin{document}

\makefrontmatter

\section{Introduction}

Chemical evolution connects the early production of the light elements in
big-bang nucleosynthesis (BBN) to the multitude of elements observed in the
Universe today.  In fact it is a crucial step in extracting the primordial
abundances of the light elements from present day observations in order to
test BBN (Walker \etal~1991; Copi, Schramm, \& Turner~1995a).  Models of
\chemevil\ have been studied in many ways since the pioneering work of Cameron
\& Truran~(1971), Talbot \& Arnett~(1971), and Tinsley~(1972, 1980).  More
recently Timmes, Woosley, \& Weaver~(1995, hereafter TWW) have performed
detailed calculations of 76 stable isotopes for one particular infall model
employing only two free parameters in their model.  Complementary to this,
Fields~(1996) explored 1460 possible \chemevil\ scenarios within the context
of a \chemevil\ framework.  This work focussed on the effects of these
\chemevil\ models on the evolution of the light elements.  Tosi~(1988)
performed a similar comparison for a number of \chemevil\ models focusing on
the heavy elements and other constraints.

All of these studies considered \chemevil\ via the standard approach; write
down the integro-differential equations that specifies the evolution of the
elements and solve them for the mean behavior expected.  However the large
sample of stars observed by Edvardsson \etal~(1993) has once again highlighted
the fact that abundances are not uniform in the solar neighborhood.  There is
an intrinsic scatter in the observed abundances as a function of time and
metallicity.  Indeed it is not surprising that this is the case.  Many
physical processes can lead to abundance differences in the solar
neighborhood.  Furthermore it is well known that the standard approach to
\chemevil\ does not attempt to address the scatter in the observations but
instead works to reproduce the average behavior.  To accurately model the
\chemevil\ of the Galaxy would require a coupling of hydrodynamics with star
formation, stellar evolution, and galactic evolution.  Besides being
computationally prohibitive, the physics of many of the processes involved is
not yet adequately understood.  Thus some assumptions and simplifications
enter into all models of \chemevil.

Numerous attempts have been made to explain the observed abundance spreads and
we will not review them in detail.  See van den Hoek \& de Jong~(1996) for
such a discussion.  These attempts range from stellar orbit diffusion coupled
with a Galactic radial abundance gradient (see e.g., Fran\c{c}ois \&
Matteucci~1993; Wielen, Fuchs, \& Dettbarn~1996) to processes that would lead
to abundance inhomogeneities from a homogeneous starting point such as
chemical fractionation in grain formation (see e.g., Henning \&
G\"urtler~1986).  The models of van den Hoek \& de Jong~(1996) consider
sequential enrichment by successive generations of stars within individual gas
clouds.  Their work bears the closest resemblance to the work discussed here.

In the work reported here we construct standard \chemevil\ models with
standard sets of parameters but solve them in a Monte Carlo fashion.  Copi,
Schramm, \& Turner~(1995b) followed a similar procedure.  In their work they
focussed on D and \He{3} and treated \chemevil\ in a parametric fashion.
Distributions for \He{3} destruction and production based on \chemevil\ models
were employed.  The benefit of this approach is that it allows the evolution
to be run backwards; something not possible in standard \chemevil\ models.
Starting from the pre-solar D and \He{3} observations the distribution of
primordial D and \He{3} abundances can be generated.  Such a distribution can
then be used to constrain BBN (Copi, Schramm, \& Turner~1995c).  Unfortunately
it is difficult to compare the results from parameterized models with standard
\chemevil\ models since the many assumptions and approximations are convolved
into the distributions chosen.

Solving standard \chemevil\ models in a Monte Carlo fashion leads to
randomness because of the different histories the material can experience as a
region evolves.  Two otherwise identical regions can end up with different
abundances due to the different numbers and types of stars formed during their
evolution.  This randomness introduced into otherwise standard \chemevil\
models allows us to study the expected spread in abundances for a particular
\chemevil\ model, not just to compare different \chemevil\ models.  In this
work we consider three models for the solar neighborhood.  We select fairly
standard, one zone models that have been well studied by other workers in this
field.  We will focus on the scatter in abundances predicted from them.  We do
not make a distinction between the halo and disk phases of Galactic evolution
in this work.  The three models considered here are a closed box model, an
infall model, and an outflow model.  The details for these models are given in
section 2.

In these models we follow 13 isotopes H, D, \He3, \He4, \Li7, \C{12}, \C{13},
\N{14}, \O{16}, \Ne{20}, \Si{28}, \S{32}, and \Fe{56}.  We find that the
scatter in the heavy, $\alpha$-elements, \O{16}, \S{28}, and \S{32}, is well
fit by the stochastic models.  The same is not true for \C{12} and \N{14}
which may indicate the need to include more physics in our prescription for
low mass stars.  The light elements exhibit very little scatter in these
models, at least for the solar neighborhood.  This is in good agreement with D
observations in the solar neighborhood but not with \He{3} observations in
\hii\ regions.  Recall that our models are tuned for the solar neighborhood. A
detailed discussion of the results can be found in section~3.  The conclusions
are given in section~4.

\section{Chemical Evolution Model}
\label{sect:models}

Here we discuss the ingredients of the chemical evolution models we will
consider throughout the rest of this work.  Since our focus is on the role of
different histories and how they affect the spread in elemental abundances we
restrict ourselves to only three, relatively simple, one zone models for the
solar neighborhood.  Furthermore, we pick a fairly standard set of parameters
for all these models.  A detailed search of parameter space for models similar
to the ones considered here can be found in the work of Fields~(1996).

\subsection{Basic Ingredients}

The main ingredient of a \chemevil\ model is the stellar birthrate function
$C(t,M)$ which gives the distribution of stars that form as a function of time
and mass.  Since a complete theory of star formation from a gas cloud is
lacking it is customary to assume that this function is separable
\be C(t,M) = \psi(t) \phi(M). \ee
Here $\psi(t)$ is the star formation rate~(SFR) and is assumed to be
independent of mass.  Similarly, $\phi(M)$ is the initial mass function~(IMF)
and is assumed to be independent of time.  We will follow a Schmidt~(1959,
1963) law for the SFR 
\be \psi(t) = \nu \sigma_{\rm Tot}(t) \left[ \frac{\sigma_{\rm
gas}(t)}{\sigma_{\rm Tot} (t)} \right]^\alpha, \label{eqn:SFR} \ee
where $\nu$ is a dimensionless parameter and $\sigma$ is the surface mass
density.  In this work we will only consider $\alpha=1$ so that $\psi(t)
= \nu \sigma_{\rm gas} (t)$.  We assume the IMF follows  the Salpeter~(1955)
form
\be \phi(M) \propto M^{-x}. \label{eqn:IMF} \ee
A power law is particularly sensitive to the
limits we place on it.  Here we use $M\in [0.08, 40] \Msun$.  The upper limit
is based on the set of high mass stellar yields employed (see
section~\ref{sect:stellar-yields}).  We will restrict ourselves to $x=2.35$
for all models except the outflow model.

We further assume that the mass of a star and its initial composition are
sufficient to describe all of its properties.  Other parameters, such as
angular momentum, could play an important role in defining the properties of
stars but are not considered here.  Due to the many uncertainties involved
even in this simplified picture, we do not employ more complicated, albeit
more realistic, stellar models.  In the \chemevil\ models considered here we
do not employ the instantaneous recycling approximation.  Instead we delay the
release of the ejecta from stars until their death as given by their lifetime.
For all stars we employ the (metallicity independent) fit of Scalo~(1986)
\be \log_{10}\tau(M) = 10.0 - 3.6\log_{10}M + \left( \log_{10} M \right)^2, \ee
where $\tau(M)$ is the stellar lifetime given in years.

\subsection{Stellar Yields}\label{sect:stellar-yields}

Perhaps the most important ingredient in a \chemevil\ model is the elemental
yields ejected from stars.  Here our assumption that stellar properties are
only a function of mass and initial composition is most evident.  Although
stellar yields calculations continue to improve, there are still numerous
assumptions present in all calculations that make predicted yields model
dependent.  The yields used in this work are discussed below.  Note that we
assume all D is burned in all stars so that the ejected abundance of D is
always zero.

\subsubsection{High Mass Stars}

For high mass stars we use the yields of Woosley \& Weaver~(1995).  The
distinction between low mass and high mass is model dependent.  According to
the models of Woosley \& Weaver all stars with mass $M>11\Msun$ form type~II
supernova and hence undergo explosive nucleosynthesis.  These models were
chosen because they are particularly meticulous, covering a fine mass grid and
five metallicities from $Z=0$ to $Z=Z_\odot$.  However, these models only
consider stars up to $40\Msun$, in part because they do not include mass loss
which can be important for higher mass, higher metallicity stars (see e.g.,
Maeder~1992, 1993; Woosley, Langer, \& Weaver~1993, 1995).  This is the origin
of the upper mass limit $M \le 40\Msun$ on the IMF~(\ref{eqn:IMF}).  Even in
these models there is considerable scatter in the predicted yields as a
function of mass (for fixed $Z$) and as a function of metallicity (for fixed
$M$).  For this reason we fit the yields as a function of mass and metallicity
so that the final results are not sensitive to these fluctuations.

	Finally, the iron yield from type~II supernova models is very
sensitive to a number of assumptions, in particular the neutron star mass cut.
Thus the iron yield is very uncertain.  Here we choose to decrease the yield
given in Woosley \& Weaver by a factor of two as suggested by TWW since this
yield appears to give a better fit to the data for a wide range of elements
(see their figure~11).

\subsubsection{Low Mass Stars}

In the standard case for low mass stars ($M<8\Msun$) we use the yields of
Renzini \& Voli~(1981).  Unfortunately these table are sparse in both mass and
metallicity.  For the \He3\ yield we use the result of Iben \& Truran~(1978).
It is well known that this yield leads to a large production of \He3\ in low
mass stars in good agreement with observations of planetary nebulae (Rood,
Bania, \& Wilson~1992; Rood \etal~1995).

More recently Hogan~(1995) has suggested that the extra mixing mechanism
invoked to explain the observed $\C{12}/\C{13}$ ratio in low mass stars (see
e.g., Dearborn, Eggleton, \& Schramm~1976) will also destroy \He3.  A number
of models have been built around this proposal; an artificial ``elevator''
type mixing (Wasserburg, Boothroyd, \& Sackmann~1995), the mixing modeled as
a diffusion process (Denissenkov \& Weiss~1996; Weiss, Wagenhuber, \&
Denissenkov~1996), and a rotationally induced mixing model (Charbonnel~1994,
1995; Forestini \& Charbonnel~1996).  Also Cumming \& Haxton~(1996) have
suggested a salt-finger like instability that could explain the solar neutrino
problem and also lead to \He3\ destruction.  Calculations of stellar yields
including this new extra mixing are on going.  For yields that include this
extra mixing we employ the work of Boothroyd \& Sackmann~(1996) which followed
the yields through second dredge up.  Note that third dredge up and hot bottom
burning are not included in this calculation.  Standard stellar models predict
that \C{13} and \N{14} should experience only minor changes to their
abundances due to third dredge up but the \N{14} yield could be enhanced in
stars with masses $M\gtsim 4\Msun$ due to hot bottom burning (Boothroyd,
Sackmann, \& Wasserburg~1995).  For stars with masses $M\ltsim2\Msun$ we
include the preliminary calculations of cool bottom processing by
Boothroyd~(1996, private communication).  The destruction of \He3\ is an
important effect that comes from this extra mixing. We include the most
extreme destruction model employed by Boothroyd \& Malaney~(1996).  Since
these models lead to net \He3\ destruction they cannot explain the
observations of planetary nebulae (Galli \etal~1996).  Thus we allow 60\% of
the models to follow these extra processing yields and 40\% to follow the
older yields of Renzini \& Voli.  A distribution of yields may be expected if
angular momentum is an important parameter in determining the mixing
experienced in stars (see also Olive \etal~1996).

	Nuclear physics could also lead to extra destruction of \He3\
if there were a low energy resonance in the $\He3+\He3$ reaction (Galli
\etal~1994). However, any such mechanism is a global effect that is not
dependent on the physical state of the star.  Thus the observation of a
large \He3\ abundance in planetary nebulae rules out such a mechanism (Galli
\etal~1996) unless there is another method of producing \He3\ in planetary
nebulae or the deduced abundances are in error.  We will not consider this
option further.

Finally we note that van den Hoek \& Groenewegen~(1996) have recently
calculated a fine grid of low mass stellar models.  They have followed the
evolution through third dredge up and include hot bottom burning.  These
yields became available after the calculations reported here were completed.
Thus we have not included them in this work.

\subsubsection{Intermediate Mass Stars}

	We are left with the uncertainty of how to treat stars in the mass
range $8 < M/\Msun < 11$ which I label as intermediate mass stars.  These
stars, under some circumstances, may undergo explosive nucleosynthesis.  On
the other hand they may eject mostly \He4\ (Woosley \& Weaver~1986) if their
core only undergoes helium burning.  To simplify the calculation and to smooth
over the sharp mass cutoffs we interpolate between the two sets of yields for
all stars in this mass range.

\subsection{Type Ia Supernovae}

Type Ia supernovae are important ingredients in any \chemevil\ model since they
produce roughly half the iron in the Universe (the exact number is model
dependent and can range from about one-third to two-thirds).  Type Ia
supernovae are the only supernovae expected to come from low mass star
progenitors.  The exact progenitors are still uncertain though they invariably
involve binary star accretion.  We follow the standard prescription of Greggio
\& Renzini~(1983) to determine the rate of type Ia supernovae.  For the yields
we employ the ubiquitous W7 model of Nomoto, Thielemann, \& Yokoi~(1984).  The
actual yields for this model and the W70 model (the zero metallicity version)
comes from the recent calculations of Nomoto \etal~(1996).

Implicitly we are assuming that all type Ia supernovae are the same, except
for the slight metallicity dependence in their yields.  The debate between the
progenitors, Chandrasekhar versus sub-Chandrasekhar mass white dwarfs, and
their evolutionary scenario, doubly degenerate versus singly degenerate, is
still on going.  In fact, more than one type of progenitor or evolutionary
scenario may be experienced in nature.  These different type Ia supernova
scenarios can lead to different yields from the explosion (Nomoto \etal~1996).
Fortunately, iron, the main product from type Ia supernovae, is relatively
insensitive to the scenario employed, though, it can be decreased by almost
40\% in some speculative models.  Other elements ejected from type Ia
supernovae, such as \C{12}, are more sensitive to the scenario.  The ejected
mass of these elements is at a much lower level than that for iron, thus the
many other uncertainties in \chemevil\ models currently precludes us from
determining the appropriate type Ia supernova scenario from the observed
abundance trends.

\subsection{Infall}

A very common ingredient to include in a \chemevil\ model is infall.  We will
include the infall of primordial material via an exponential infall rate, \be
I = I_0 e^{-t/\tau_{\rm inf}}\; [\Msun\Gyr^{-1}]. \label{eqn:infall} \ee 
Here $\tau_{\rm inf}$ is
the characteristic infall time and $I_0$ is a normalization constant.  The
constant $I_0$ is determined from the total amount of material that infalls
and the total evolutionary time.  Note that here we only consider the simple
case of 90\% of the material coming from primordial infall with $\tau_{\rm
inf} = 5\Gyr$.

\subsection{Outflow}

The last extra ingredient we will consider here is outflow.  For outflow we
allow both type Ia and type II supernovae to force some fraction of their
ejecta out of the region into the intergalactic medium~(IGM) during the
explosion.  We do not consider the fact that supernovae can also heat the
interstellar medium~(ISM) driving some of this material into the IGM such as
is considered in the models of Scully \etal~(1996).  The correct prescription
for including such heating is not well understood.  Here we will consider the
case where 65\% of the ejecta from both type Ia supernovae and type II
supernovae is blown from the region.

There are many other options for including outflow.  We do not consider models
that preferentially blow out metals but leave the lower mass elements behind
(see e.g., Copi, Schramm, \& Turner~1995b).  In such a model the light
elements are blown from the surface of the star in a wind prior to the
supernova explosion which creates and ejects the heavy elements.  Similarly we
could consider a merger model of galaxy formation (Mathews \& Schramm~1993)
where low mass objects merge to form the Galaxy.  The \chemevil\ of these low
mass regions would be susceptible to outflow due to their low escape velocity.
In such a model outflow is an important ingredient and could be at a much
higher level than considered here.  Indeed a large outflow may be a necessary
feature of \chemevil\ models.  Observations of hot gas in clusters shows that
the intercluster medium is enriched in metals consistent with type II
supernova trends (Fukazawa \etal~1996; Mushotzky \etal~1996).  Some early work
on cluster \chemevil\ has been performed (Lowewnstein \& Mushotzky~1996;
Matteucci \& Gibson~1996).  More detailed models will benefit from the on going
observations that continue to enlarge and improve the data set.

\subsection{Stochastic Models}

The standard approach for solving a chemical evolution model is to write down
the integro-differential equation that describes the flow of gas into and out
of stars.  This equation is then solved numerically to obtain the time
evolution of elemental abundances and other properties of the system (see
e.g., Tinsley~1980).  This approach has been followed extensively in the past
and provides good results on the average behavior of the properties studied
(see e.g., Fran\c{c}ois, Vangioni-Flam, \& Audouze~1990; Steigman \&
Tosi~1995; TWW; Fields~1996; and references therein).

\subsubsection{Constructing a History}

To probe the distribution of abundances expected we solve the problem in a
Monte Carlo fashion.  We start with a gas cloud of some total mass $M_{\rm
Tot}$.  At each time, $t$, we want to know what happens to the region over the
time interval $\delta t$.  To begin, stars will form from the available gas.
The number of stars, $N_\star$, that form is a random number that on average
follows the SFR~(\ref{eqn:SFR}) with a mean number of stars formed
\be \bar N_\star = \nu \frac{M_{\rm gas}}{\bar M} \delta t. \ee
Here $\bar M$ is the average mass of a star determined from the
IMF~(\ref{eqn:IMF}) and we have explicitly assumed $\alpha=1$ for the SFR\@.
The number of stars formed is then drawn from a Poisson distribution 
\be P(N_\star) = \frac{\bar N_\star^{N_\star}}{N_\star!} e^{-\bar
N_\star}. \ee
For each new star we randomly pick its mass from the IMF~(\ref{eqn:IMF}). Once
we know its mass we know its lifetime and abundance yields.  All stars created
at time $t$ start with the abundance of the gas they were created from.  For
each star created we remove its mass from the available mass in the gas.  The
abundances in the gas are unchanged by stellar births.

After all the stars for the time $t$ are created we mix in material from stars
that have died during the current time interval.  Note that although we do not
use the instantaneous recycling approximation, we still assume that all ejecta
from stars are instantaneously mixed in the region that we are evolving.
Clearly the finite mixing time is an important consideration.  However, since
we are evolving a region much smaller than the entire galaxy this
approximation is not as extreme as in the standard case.  The mass fraction of
each element, $i$, in the gas changes by 
\be X_i = \frac{X_i^0 M_{\rm gas} + X_i^{\rm out} M_{\rm ej}}{M_{\rm gas} +
M_{\rm ej}}, \ee
where $X_i^0$ is the original mass fraction of element $i$ in the gas,
$X_i^{\rm out}$ is the mass fraction ejected by the dying star, and $M_{\rm
ej}$ is the total mass of material ejected by the star.  The total gas mass is
then increased by $M_{\rm ej}$.  This material is now available for subsequent
star formation. At this time we also take care of any infall or outflow, both
of which affect the total mass and the gas mass in the region.  We repeat this
process until we reach $t_0$, the total evolutionary time.  This defines one
history the material in the region could have experienced.

\subsubsection{Ensemble Averages}

We have now described how to find a particular history for a region.  But
there are many possible histories for the material.  Starting from the same
initial conditions, the same SFR, IMF, low mass stellar yields, choice of
infall or outflow, etc., we construct many histories for a particular region.
Since the number and mass of the stars will be different at different time
steps for each history, the final abundances will also be different.  The
ensemble of these regions along with the initial conditions defines our
models.  The predicted spread in abundances due to different histories can then
be extracted from the many regions we have evolved. 

\subsubsection{Model Descriptions}

For all models considered here we evolve a region of total mass, $M_{\rm Tot}
= 10^5\Msun$.  This is the approximate mass of current star forming regions
such as Orion (Shields~1990).  Throughout this work we have assumed that these
regions do not mix with neighboring regions which may not be a valid
assumption for the solar neighborhood.  Furthermore, regions of this size lead
to the best fit for many of the heavy elements as discussed below.  A smaller
mass region would exhibit far more scatter and a much larger mass region would
be equivalent to solving the integro-differential equation for the mean
behavior.  By choosing this mass we are explicitly assuming a mixing scale of
$10^5\Msun$.  For the solar neighborhood today this corresponds to a region of
radius $\sim 100\unit{pc}$.  Typically models of the solar neighborhood assume
that the whole region is well mixed corresponding to a mass scale
$10^8\hbox{--}10^9\Msun$.  Thus we are assuming mixing on a much smaller scale
than typically employed.  Furthermore, observations of D in the local ISM
(Linsky \etal~1993, 1995) find identical abundances along different lines of
sight.  These observations argue that material is well mixed on scales of
about $10^4\Msun$; only an order of magnitude smaller than we have assumed in
this work.  Finally, the velocity required to travel from one edge of the
region to the other in the time $\delta t \sim 10^6\yr$ is $v \sim 100\kmps$,
a reasonable value for supernova ejecta.  Thus although we are assuming the
region is well mixed on this mass scale, it is a reasonable assumption and not
as demanding as assuming the entire solar neighborhood is well mixed.

We assume the age of the galaxy is $t_0=15\Gyr$ and the age of the sun is
5\Gyr.  The constant $\nu$ in the SFR~(\ref{eqn:SFR}) is tuned to get the
present day gas-to-total mass fraction, $\mu$, in the range 5\%--20\%
(Rana~1991).  In fact, all models have $\mu\sim12\%$ and the results are
fairly insensitive to the final value of $\mu$.  The type Ia supernova rate is
tuned to get the solar iron abundance correct (Anders \& Grevesse~1989).  All
other parameters are fixed a priori.

The models we consider are discussed here.  The first is the standard closed
box model with no infall and no outflow.  The second is an infall model.  Here
we allow 90\% of the material to be primordial infall~(\ref{eqn:infall}) with
a characteristic time scale of $\tau_{\rm inf} = 5\Gyr$.  The final is an
outflow model.  Here we allow 65\% of the ejecta from both type Ia and type II
supernovae to escape the region.  In this model the region starts with $M_{\rm
Tot}$ in gas and typically ends with 80\% of the material left in the region.
Thus the mass of material ejected into the IGM is about twice the total
mass in gas left in the region.  In this model we also flatten the slope of
the IMF~(\ref{eqn:IMF}) to $x=2.1$ to take into account the extra processing
allowed by the loss of material to the IGM\@.  A different strategy that
allows for even more processing is to introduce a time dependent IMF that is
skewed towards high mass stars at early times (Scully \etal~1996; Olive
\etal~1996).  We do not consider this options here.

The initial abundances for all models are taken from standard, homogeneous BBN
(see e.g., Copi, Schramm, \& Turner~1995a).  Only the light elements D, \He3,
\He4, and \Li7\ are created in BBN\@.  All other initial abundances are
assumed to be zero.  We allow for two values of the baryon-to-photon ratio,
$\eta$; $\eta=4.5\sci{-10}$ and $\eta=5.5\sci{-10}$.  The higher value of
$\eta$ is consistent with the low deuterium observations in two quasar
absorption systems (Tytler, Fann, \& Burles~1996; Burles \& Tytler~1996) if we
interpret them as primordial.  We do not consider $\eta\approx2\sci{-10}$
which is necessary to explain the high deuterium observation (Carswell
\etal~1994; Songaila \etal~1994; Rugers \& Hogan~1996).  Olive \etal~(1996)
have constructed an outflow model with a time dependent IMF that can fit this
observation.  Since we do not include a time dependent IMF we do not consider
models with such high primordial deuterium.  Finally we have calculated the D 
and \He{3} evolution for a model with $\eta = 3.2\sci{-10}$ to show that
higher values of $\eta$ are allowed even in the simple models we construct
here.  A detailed study of this model will not be included in this work.

Finally, as discussed above, we allow for two options with low mass stars.
Either we employ the standard yields of Renzini \& Voli~(1981) coupled with
\He3\ production given by Iben \& Truran~(1978) or we employ the models with
extra mixing and \He3\ destruction as implemented by Boothroyd \&
Sackmann~(1996).  Thus for each of the three models we have four sets of
parameters; two for the low mass star options and two for the initial
abundances.

\section{Results}

For each model we have evolved 1000 regions to determine the expected spread
in abundances.  In all the results discussed here we will quote the ranges in
which 95\% of the models fall.  In general since we are considering fairly
standard \chemevil\ models the average behavior of our models is in good
agreement with previous work (see e.g., TWW; Fields~1996).  We will focus on
the distribution of abundances produced since this is the new feature of the
work reported here.  Recall that we generate every star that is created as a
region of gas evolves.  We find that roughly $4\sci{5}$ stars are formed per
region in all three models.

\putfig{age-metallicity}

\subsection{Age-Metallicity Relation}

The age-metallicity relation expected for the three models is shown in
figure~\ref{fig:age-metallicity}.  There is little difference among the
predicted age-metallicity relations for the models we consider.  We immediately
see that differences in the history alone are not sufficient to explain the
spread in observed abundances.  The predicted spread is about 0.2\dex\ whereas
the observed spread is about 1\dex.  There are a number of difficulties in
making this comparison between theory and observation.  The age of a star is
not an observed quantity but is instead deduced from isochrone fitting.
Furthermore, the age of the Universe is not known precisely.  To plot the data
in figure~\ref{fig:age-metallicity} we assumed an age for the Universe of
15\Gyr.  The fact that the observed age-metallicity relation does not appear
to decrease rapidly at early times as expected based on an initial zero
metallicity Universe can be traced to this fact.  The best fit isochrones for
some stars have an age greater than 15\Gyr, albeit with a large uncertainty.
Models with a prompt initial enrichment of iron can be constructed that better
reproduce the high iron abundances at early times.  For example, a model with
an IMF skewed toward high mass stars at early times would lead to this type of
enrichment.  Though the data allows for this type of enrichment they don't
require it.

However, the uncertainties in the age alone are not sufficient to explain the
discrepancy in the predicted and observed spreads.  Since the iron abundance
remains relatively flat for most of the history of the Universe, extremely
large errors would be necessary to be consistent with the predictions.
Immediately we find a shortcoming with the approach we are employing.  It
cannot explain the observed spread in the age-metallicity relation.  The
reasons for this are unclear, but may point to the need for extra physics that
is not included in the current models such as multiple evolutionary scenarios
for type Ia supernovae.  It is interesting to note that the age-metallicity
relation plays an entirely different role in constraining stochastic models
than it does for standard models.  In the standard case the large spread in
the observations means that almost any model is consistent with the
observations.  Here the large scatter points to a shortcoming of the model
that must be corrected in order to accurately reproduce the observed spread.

\subsection{Heavy Elements}

The heavy $\alpha$-elements we consider in this work are \O{16}, \Ne{20},
\Si{28}, and \S{32}, all of which are predominantly made in type II
supernovae.  Since the heavy elements do not depend on our choice of $\eta$,
they always start at zero abundance, nor the low mass yields, the results
discussed here are a global features of each model.  Shown in
\putfig{heavy-solar}
figure~\ref{fig:heavy-solar} are the results for our three models at the time
of the formation of the sun (we have assumed the age of the Universe is
10\Gyr\ at this time).  We have plotted the ratio of the predicted mass
fraction to the solar value.  A ratio of 1 indicates perfect agreement with
the observed solar value.  As we found in the discussion of our \chemevil\
models (section~\ref{sect:models}) there are many uncertainties and
assumptions that go into constructing such models.  In particular the stellar
yields are uncertain.  Thus perfect agreement is not a reasonable expectation.
Due to the difficulty in assessing all of the uncertainties introduced into
our models and in the observations we allow ourselves a factor of two range
in comparing the observations with the predictions (TWW).  As we
can see, for all three models we find good agreement between the models and
the observations, though, the predictions for \Ne{20} are somewhat low.
Recall that \Fe{56} was used to tune the type Ia supernova rate so it is not
surprising that it agrees well with the observations.  The predicted spread
for these elements has an interesting size and we will focus on it now.

\putfig{O-Fe}
\putfig{Ne-Fe}
\putfig{Si-Fe}
\putfig{S-Fe}

	To better study the predicted spread in abundances and since the age
of the Universe when an individual star is formed is not a directly measurable
quantity we follow the convention of plotting our results as a function of the
iron abundance which is directly measured in each star.  Shown in
figures~\ref{fig:O-Fe}--\ref{fig:S-Fe} are the results along with a
representative sample of observations.  See TWW for a detailed discussion of
the observations for each element.  As we noted above, the overall
normalization of the curves is somewhat uncertain so we do not expect them to
perfectly overlay the data.  In the three cases where observations are
available the predicted spread in abundances is in good agreement with the
observed spread, particularly in the region of iron abundance $[\feh] \gtsim
-1$.  Thus a $10^5\Msun$ region does a good job of explaining the observations
based solely on the different evolutionary history that these regions undergo.
The raggedness in these plots for $[\feh] \ltsim -1$ is due to the fact that
the iron abundance climbs to nearly solar on a time scale of about 1\Gyr\ (see
figure~\ref{fig:age-metallicity}).  Thus the statistics for the lower iron
abundances are poor.  However the scatter at low iron abundances offers some
hope of distinguishing between different types of \chemevil\ models.  As shown
in the figures, the infall model predicts large scatter at low iron abundances
since there is very little material in the region at early times.  Further
observations similar to those by McWillian \etal~(1995) and Ryan, Norris, \&
Beers~(1996) for $[\feh] \ltsim -2$ will help clarify the situation.

\subsection{Carbon and Nitrogen}

Each of \C{12}, \C{13}, and \N{14} are sensitive to the choices we make
regarding low mass stellar yields.  They further involve other complications
that make their interpretation difficult.  We will discuss each in turn.  As
with the heavy elements, carbon and nitrogen are not made in BBN, thus the
results are independent of the value of $\eta$ chosen.

\subsubsection{Carbon}

Carbon-12 is made in a wide range of stars; whereas carbon-13 is produced
mainly in low mass stars and is very sensitive to processing in these stars.
Cool bottom processing strongly affects the final \C{13} abundance.  Not all
low mass stars are the same.  The evolutionary history of $1\Msun$ and
$5\Msun$ stars are quite different.  Furthermore the mixing history of the
star can radically change the final yields of \C{12}\ and \C{13}.  In fact,
the low number ratio of $\C{12}/\C{13}$ observed in the envelopes of low mass
stars was an early motivation for considering an extra mixing mechanism in low
mass stars (Dearborn, Eggleton, \& Schramm~1976).

\putfig{CN-solar}

Shown in figure~\ref{fig:CN-solar} are the predicted ratios of \C{12} and
\C{13}, to the observed solar values.  Also shown is a comparison to the solar
$\C{12}/\C{13}$ value.  As expected, carbon is quite sensitive to our choice
of low mass stellar yields.  For \C{12} the extra mixing models produce
somewhat less \C{12} than the older yields; although both sets are within our
factor of two uncertainty.  In contrast, \C{13} is very strongly affected by
mixing in low mass stars.  The predicted \C{13} abundance changes from being
about one half solar with a relatively small predicted range from the older
yields to being greater than twice solar value with a very large predicted
range from the newer yields.  In fact, the outflow model produces no histories
within the allowed factor of two uncertainty.  Furthermore the very large
spread is due, in part, to the mixture of old and new low mass stellar yields.
Since the \C{13} yields are quite different between the two calculations we
end up with a wide range of final \C{13} abundances. We must keep in mind that
this difference is due largely to the preliminary cool bottom processing
yields employed.  The magnitude of the difference is likely to change as the
calculations are refined, though, the general character of the difference
should remain.

The \C{12}/\C{13} ratio suffers from this behavior in \C{13}.  Since the low
mass stellar yield of \C{13} is not well understood we cannot hope to learn
much from this ratio.  We note that the \C{12}/\C{13} ratio varies from about
three times solar from the older yields to about one half solar for the newer
yields.  Again due to the preliminary nature of the cool bottom processing
yields it is premature to draw strong conclusions from these results.

\putfig{C-Fe}

The evolution of carbon relative to iron is shown in figure~\ref{fig:C-Fe}.
Immediately we see that unlike the heavy elements (see e.g.,
figure~\ref{fig:O-Fe}) the agreement between the models and observations is
not very good.  Again the mean abundance can be shifted somewhat due to
uncertainties in the stellar yields.  The data shows significantly more
scatter than we saw in the heavy elements.  In fact, there is no obvious trend
in the data.  The large spread in the data indicates that other factors are
important in determining the carbon abundance.  In particular the rotational
history of stars may help explain the scatter.  If meridional mixing in stars
is sensitive to their rotational history and produces extra mixing in stars
then a distribution of angular momenta in these stars could lead to a spread
in the carbon abundance.  The difference in histories only produces about half
of the observed scatter.  Since detailed models of this type are not currently
available we will not pursue this possibility here.  Although suggestive, we
must also keep in mind that low mass stars are difficult to evolve.  They
experience multiple dredge up events and thermal pulses.  Furthermore they are
sensitive to the depth and type of convection that occurs.  Thus it is
premature to claim understanding of their stellar yields or their affects on
\chemevil.

\subsubsection{Nitrogen}

Besides the sensitivity to stellar models discussed above, nitrogen has the
added difficulty that it can be created as both a primary and a secondary
element in stellar nucleosynthesis.  Frequently stellar models only include
the secondary production from carbon seeds.  Here we directly employ the
yields given without considering the question of primary versus secondary
production.  Due to this uncertainty in the production of nitrogen we cannot
use it to study \chemevil\ but instead can use \chemevil\ to learn about
nitrogen production (see Fuller, Boyd, \& Kallen~1991; Fields~1996).  An
understanding of the evolution of nitrogen is important since nitrogen is
frequently used as a tracer for metallicity when determining the primordial
\He4\ abundance (Olive \& Steigman~1995; Olive \& Scully~1996).  The
functional form used to extrapolate to zero metallicity changes depending on
the mixture of primary and secondary nitrogen.

The solar ratio of \N{14} for all models is quite low
(figure~\ref{fig:CN-solar}).  Since we do not include hot bottom burning in
the newer yields it is not surprising that there is little difference
between the two sets of yields.  The overall magnitude of the \N{14} abundance
is expected to be affected by both this hot bottom burning and by the choice
of primary versus secondary production.

\putfig{N-Fe}

Shown in figure~\ref{fig:N-Fe} is the nitrogen abundance as a function of the
iron abundance.  The results here are quite similar to those for carbon
(figure~\ref{fig:C-Fe}).  A similar discussion applies.  Again only about a
half of the observed spread can be explained by the different histories.
These results may be indicative of the necessity to include an extra
parameter, such as angular momentum, when describing stars in \chemevil\
models.

\subsection{Lithium-7}

Lithium-7 is an important element since it is the heaviest one produced in
measurable quantities in the big-bang.  \Li7\ is made in the $\nu$-process in
type II supernovae (Woosley \& Weaver~1995).  The exact \Li7\ abundance is
sensitive to the choice of the $\mu$ and $\tau$ neutrino temperatures.  We
have included the $\nu$-process from the calculations of Woosley \& Weaver but
note that these yields could still be quite uncertain.  Furthermore the
evolution of \Li{7} is complicated by the fact that it is also created in
cosmic ray nucleosynthesis (Walker \etal~1993).  Figure~\ref{fig:lihe-solar}
\putfig{lihe-solar}
shows the ratio of the predicted abundance to the observed solar abundance.
Note that \Li7\ is sensitive to the choice of $\eta$, which changes the
initial conditions, and the low mass stellar yields we employ.  In general we
find about half the solar \Li7\ can be accounted for with the yields employed
here.  The origin of the rest of the \Li{7} is still uncertain.  Cosmic rays
can only produce about 10\% of the predicted \Li{7} (Vangioni-Flam \etal~1996)
thus they cannot account for this deficit.  This discrepancy is not too
worrisome, though, since the \Li{7} yield from the $\nu$-process in type II
supernovae is still uncertain by at least a factor of two.

\putfig{Li}

In figure~\ref{fig:Li} we show the evolution of $\log_{10} N({\rm Li}) \equiv
12 + \log_{10} ({\rm Li/H})$ as a function of the iron abundance.  The \Li7\
values with an iron abundance $[\feh] \ltsim -2$ show the Spite plateau (Spite
\& Spite~1982) that is used to determine the primordial \Li7\ abundance.  The
discrepency between the primordial value and the Spite plateau is not
unexpected.  It either argues for a lower value for $\eta$ or for some \Li{7}
depletion in stars.  Many stellar models predict about a factor of two
depletion of \Li{7} in stars observed on the Spite plateau (see e.g.,
Pinsonneault, Deliyannis, \& Demarque~1992; Chaboyer \& Demarque~1994;
Vauclair \& Charbonnel~1995).  To be consistent with the higher primordial
starting value, $\eta = 5.5\sci{-10}$, requires a factor of 3--4 depletion.
Such a level of depletion seems inconsistent with the observations (Lemoine
\etal~1996).

The models do a good job of explaining the general trend of the data.  We
expect them to serve as an upper envelope to the observations with an iron
abundance $[\feh] \gtsim -1$ due to \Li7\ destruction in stars.  The scatter
in the \Li7\ abundance, particularly on the Spite plateau, is quite small.
Some of the scatter is due to systematic uncertainties in the stellar
atmosphere models emplyed to convert the observed line strength into an
abundance.  The extremely small spread predicted here shows that different
evolutionary histories do not play a role in defining an intrinsic spread in
the Spite plateau.  If intrinsic scatter does exist in the Spite plateau it
must be explained via stellar processing (see e.g., Vauclair \&
Charbonnel~1995) not by \chemevil.  In fact, this shows that \chemevil\ does
not introduce an intrinsic scatter at a level that would be difficult to
extract from scatter introduced by stellar processing.

\subsection{Light Elements}

The \chemevil\ of \D\ and \He4\ is easy; stars make \He4\ and stars destroy
all of their \D\ during their pre-main sequence evolution.  The \chemevil\ of
\He3\ is more complicated.  Thus we allow for two different types of \He3\
evolution as previously discussed.  Since \D, \He3, and \He4\ are all made in
appreciable quantities in BBN, their abundance histories are sensitive to the
choice of $\eta$ which determines their initial abundances.

\subsubsection{Helium-4}

The solar abundance ratio of \He4\ is shown in figure~\ref{fig:lihe-solar}.
Since the predicted solar abundance of \He4\ is insensitive to both the
stellar yields and the choice of $\eta$, only one range for each model is
shown in the figure.  We see that there is very good agreement with the solar
value, that it is largely independent of the \chemevil\ model, and the scatter
is extremely small.  This is not surprising.  BBN production provides a large
initial abundance for \He{4}, $Y_{\rm BBN}\approx 0.24$, and is only
logarithmically dependent on the choice of $\eta$ (Walker \etal~1991).  The
solar value of \He4, $Y_\odot=0.275$ (Anders \& Grevesse~1989), is very close
to the primordial value.  Even without any production of \He4\ we find a ratio
$Y_{\rm BBN}/Y_\odot \approx 0.9$ which is well within the expected
uncertainty.  Any production of \He4\ serves to improve this agreement.  Since
only a small fraction of the final \He4\ is produced in stars and since all
stars make \He4, we expect the scatter to be quite small as is observed.

\putfig{Y-O}

In figure~\ref{fig:Y-O} we show the \He4\ abundance as a function of ${\rm
O/H}$.  The most precise observations come from extra-galactic \hii\ regions.
The data shown in the figure is a representative sample from Pagel
\etal~(1992).  A more complete sample can be found in Olive \& Scully~(1996).
As noted above we do not consider any special options for \N{14} production so
we will not discuss the behavior of \He4\ as a function of \N{14}.  See
Fields~(1996) for a thorough discussion.  The fact that the initial value in
these models is high compared to the data is well known and may be due to a
systematic shift required in the data (Copi, Schramm, \& Turner~1995a).
Alternatively a model with $\eta\approx 2\sci{-10}$ would allow the initial
value of \He4\ to be in good agreement with the data as plotted.  Note that a
linear relation is predicted as is expected and commonly employed (Olive \&
Steigman~1995).  The shallow slope of the line is not in good agreement with
the observations.  This is a standard failing of \chemevil\ models.
It may be due to our lack of knowledge regarding intermediate mass stars,
$8< M/\Msun< 11$.  Recall that we have interpolated between our high mass and
low mass tables for these stars.  If the suggestion of Woosley \&
Weaver~(1986) that these stars return mostly \He{4} to the ISM is employed
the slope steepens as expected (Fields~1996).

\subsubsection{Deuterium and Helium-3}

\putfig{D-He-v-t}

The \chemevil\ of \D\ and \He3\ are closely connected; \D\ is burned to \He3\
during the pre-main sequence of all stars.  The time evolution of \D, \He3,
and $\D + \He3$ is shown in figure~\ref{fig:D-He-v-t}.  The tightest
constraint comes from the ISM abundance of \D\ since it is precisely measured
(Linsky \etal~1993, 1995).  Note that D is not directly measured in the sun,
consistent with our assumption that all D is burned to \He{3} in its pre-main
sequence evolution.  Instead the D abundance is deduced from \He{3} in the
solar wind, which we assume is the ${\rm D} +\He3$ the sun started with (Geiss
\& Reeves~1972), and \He{3} observed in gas rich meteorites (Black~1972).
However D has been measured in the atmosphere of Jupiter via the DH/H$_2$
molecular abundance ratio (Niemann \etal~1996) which should also give a
measure of the D in the pre-solar nebula.  This value is somewhat higher than
the deduced value but the uncertainties are quite large as the ratio is
sensitive to chemical fractionation in the atmosphere.  Helium-3 has recently
been measured in the local ISM by the Ulysses satellite (Gloeckler \&
Geiss~1996).  The high redshift D observations are from quasar absorption
systems (Tytler, Fann, \& Burles~1996; Burles \& Tytler~1996).  We note that
${\rm D} + \He3$ has remained roughly constant over the past 5\Gyr (Gloeckler
\& Geiss~1996).  This has important implications for \chemevil\ models (Turner
\etal~1996) though the uncertainties are still too large to impose tight
constraints.

	The extreme production of \He3\ predicted by Iben \& Truran~(1978)
cannot be easily accommodated by any models (see
figure~\ref{fig:D-He-v-t}a,c), unless we push the already large uncertainties
on the observations to their 2- or 3-$\sigma$ values.  The infall model
(figure~\ref{fig:D-He-v-t}a,c) is the only one that approximately reproduces
the ISM D observations but is only marginally in agreement with the meteoritic
\He{3} observation.

	The models with the new low mass stellar yields do a much better job
of fitting both the meteoritic and ISM \He{3} observations (see
figure~\ref{fig:D-He-v-t}b,d).  In fact, since we employ extreme \He{3}
destruction models, it is somewhat under produced in our results.  The ISM D
is best fit by the closed box model (figure~\ref{fig:D-He-v-t}b) for $\eta =
4.5\sci{-10}$ and by the infall model (figure~\ref{fig:D-He-v-t}d) for
$\eta=5.5\sci{-10}$.

In all the models considered here there is relatively little D destruction
between the time of BBN and the observations of D in quasar absorption systems
at $z=3\hbox{--}4$, only about 10\%\@.  However, since the BBN production of D
is very sensitive to $\eta$ (Copi, Schramm, \& Turner~1995a) it is important
to account for the \chemevil\ when determining the primordial value of D\@.
If we assume the observations are the primordial abundance then it is more
consistent with $\eta \approx 6\sci{-10}$.  

\putfig{D-ja}

We have also considered the evolution of D and \He{3} in an outflow model with
$\eta = 3.2\sci{-10}$.  As is well known, \chemevil\ models can be constructed
from a wide range of initial abundances of D and \He{3} that are consistent
with the present day observations.  In figure~\ref{fig:D-ja} we show the
evolution for an outflow model where 90\% of the ejecta from type II
supernovae and 85\% of the ejecta from type Ia supernovae escapes the region.
in this model we used a 70\%/30\% mixture of the new and old low mass stellar
yields.  Furthermore we flattened the IMF to $\phi(M) \propto M^{-1.9}$ to
allow for extra processing by high mass stars.  Even more processing can be
obtained by skewing the IMF to high mass stars at early times (Olive
\etal~1996).

Finally we note that the spread in D and \He{3} is predicted to be quite
small.  For D this is in good agreement with the observations of Linsky
\etal~(1993, 1995) who found nearly identical D abundances along two different
lines of sight in the ISM\@.  This lends support to our assumption that a
$10^5\Msun$ region is well mixed in the solar neighborhood.  For \He{3} this
does not agree well with the observations in \hii\ regions which find a range
of values, $\He3/{\rm H}\approx$(1--4)\sci{-5} (Balser \etal~1994).  Note,
however, that the models discussed here are tuned for the solar neighborhood,
not \hii\ regions, and thus expecting them to reproduce these results may not
be reasonable.  The \chemevil\ model appropriate for an \hii\ region could be
quite different than the ones employed for the solar neighborhood.

\subsection{Additional Constraints}

There are a number of other constraints on galactic chemical evolution models
that we will discuss now.  We have already used the fact that the present day
gas-to-total mass fraction, $\mu=10$\%--13\% for all models, to fix the star
formation rate so this constraint is trivially satisfied.

\subsubsection{Present Day Mass Function}

\putfig{pdmf}

The present day mass function (PDMF) is the distribution of stars by mass
expected to be observed in the Universe today.  Due to different lifetimes for
stars of different masses, the PDMF is not the same as the IMF\@.
Figure~\ref{fig:pdmf} shows a comparison of the predicted PDMF and
observations from Scalo~(1986).  Only the curve for the closed box model is
shown since all models exhibit similar behavior with minor differences at low
masses, $\log M \ltsim -0.5$.  The agreement with stars above solar mass,
$\log M \gtsim 0$, is quite good.  Below this the model and observations begin
to diverge.  This is a common feature of a power law IMF; it cannot have
curvature at low mass as required by the data.  These stars do not directly
contribute to the \chemevil\ of the Universe since their lifetime is
comparable to the age of the Universe (or larger).  However, by over counting
the number of low mass stars we lock some gas into these stars that could have
gone into forming other, higher mass stars.  Though only a small amount of the
total mass is locked into these stars.

\subsubsection{G Dwarf Distribution}

\putfig{gdwarf}

The iron abundance in G-dwarf stars provides a crucial and difficult, test for
all \chemevil\ models.  The results for our models along with the data from
Rocha-Pinto \& Maciel~(1996) are shown in figure~\ref{fig:gdwarf}.  Our
results are in good agreement with those in Scully \etal~(1996).  We only show
the results for the infall and outflow models in figure~\ref{fig:gdwarf}.  The
scatter predicted in the G-dwarf distribution is fairly large.  Though it is
not sufficiently large to explain the observed distribution.  The problem
still remains that the models do not predict enough stars at $[\feh] \sim
-0.3$ and predicts too many stars at $[\feh]\sim 0$. The G-dwarf distribution
is intimately related to the age-metallicity relation.  The model results are
consistent with the fast rise in the metallicity to an almost constant value
$[\feh] \approx 0$ for most of the evolution.  To reproduce the observed
G-dwarf distribution requires slowing the rise in iron abundance.

\subsubsection{Supernova Rates}

The type II supernova rate for all of our models is approximately $2\sci{-6}$
supernovae per $10^5\Msun$ per century.  The type Ia supernova rate for all of
our models is approximately $1\sci{-7}$ supernovae per $10^5\Msun$ per
century.  If we assume the solar neighborhood is typical for the entire
Galactic disk, $M_{\rm disk} = 10^{11}\Msun$, we predict about 2 Galactic
supernovae per century.  This is in good agreement with the observed Galactic
supernova rate of $2.5^{+0.8}_{-0.5}$ per century (Tammann, L\"offler, \&
Schr\"oder~1994).

The ratio for type Ia to type II supernovae for our models is 6\% for the
closed box model, 12\% for the infall model, and 19\% for the outflow model.
The expected value is about 10\% (Tammann, L\"offler, \& Schr\"oder~1994).
The ratio is sensitive to the iron yields for both types of supernovae.  The
type Ia supernova rate in the outflow model is also quite sensitive to our
choice of the fraction of ejected material that escapes from the region.  

\section{Conclusions}

Three fairly standard, one zone \chemevil\ models have been solved for the
solar neighborhood in a stochastic manner
in order to study the expected spread in abundances due to the different
evolutionary histories the material could have undergone.  In all cases the
average behavior is in good agreement with previous work.  We have studied the
evolution of a $10^5\Msun$ region for a closed box model, an infall
model, and an outflow model of the solar neighborhood.  A region of
$10^5\Msun$ does a very good job of explaining the observed scatter as a
function of iron abundance for the heavy $\alpha$-elements, \O{16}, \Si{28},
and \S{32} (see figures~\ref{fig:O-Fe}--\ref{fig:S-Fe}).  In fact, the large
predicted spread for abundances at $[\feh] \ltsim -2$ in the infall model may
help to set limits on the amount of infall in the solar neighborhood.

In most other cases the predicted and observed spread are not in good
agreement.  This helps to point out where other physical processes may play an
important role.  The spread in the age-metallicity relation
(figure~\ref{fig:age-metallicity}) is not well fit by the predictions.  Most
of the iron in the Universe comes from type Ia supernovae in our models.  We
have only allowed for one type of evolutionary scenario for these supernovae.
The calculations of Nomoto \etal~(1996) do find a small range of iron yields
for different evolutionary scenarios for supernovae.  Though this alone is not
sufficient to explain the spread in observations. Unlike in standard
\chemevil\ models, the age-metallicity relation plays an important role in
constraining stochastic models.  A complete understanding of stochastic
\chemevil\ requires an explanation of the full spread in the age-metallicity
relation.

For carbon and nitrogen we also see that the predicted spread only accounts
for about a half of the spread found in the data (figures~\ref{fig:C-Fe}
and~\ref{fig:N-Fe}).  In this case we may be learning something about
stellar evolution.  In particular our assumption that the mass and initial
composition are sufficient to determine all properties of a star may not be
correct for low mass stars.  Unlike high mass stars, the yields of low mass
stars is dependent on many physical processes that are poorly understood and
difficult to approximate in an accurate manner.  In this work we have included
yields due to extra mixing processes in stars.  If this mixing is coupled to
the rotational history of the star we would expect a distribution of yields
for low mass stars based on a distribution of angular momenta in these stars.
Such a distribution coupled with the different evolutionary histories may
explain the scatter in the carbon and nitrogen abundance data.

Though we have included \C{13} in our work its interpretation is much more
difficult.  The evolution of \C{13} is strongly dependent on the cool bottom
processing and mixing in low mass stars.  Thus all results are very model
dependent.  We have found that the solar \C{13} abundance and the solar
\C{12}/\C{13} ratio are not well fit by any of the models
(figure~\ref{fig:CN-solar}).  The older yields predict that low mass stars
make only about half the solar \C{13} whereas the newer yields make 2--3 times
the solar abundance.  This is most evident for the case of the outflow model
where \C{13} is over produced in the new low mass stellar model.  Note that it
shows up strongly in the outflow model since there is so much extra processing
of material.  If we had included a time dependent IMF skewed towards high mass
stars at early times the behavior of \C{13} would not be as extreme.

{}From our studies we have shown that the range in baryon-to-photon ratio,
$\eta\approx (3\hbox{--}5.5)\sci{-10}$ is consistent with the present day
observations of D and \He{3} for standard \chemevil\ models.  Of course we
have only considered simple models here.  Models that allow for and even
larger range of $\eta$ can, and have been, constructed (see e.g., Olive
\etal~1996).  In general we find that the predicted spread for all the light
elements is quite small.  This further justifies the common use of \chemevil\
to extract the primordial abundances from present day observations.  Different
evolutionary histories, at least for the solar neighborhood, do not introduce
a significant amount of scatter and thus does not further complicate such
attempts.

For \Li{7} the small spread on the Spite plateau (figure~\ref{fig:Li}) argues
against an intrinsic spread in the abundances due to the different histories
the material could have gone through.  Any intrinsic scatter would instead be
due to stellar processing (Vauclair \& Charbonnel~1995).  Similarly, the
scatter in \He{4} is extremely small (figure~\ref{fig:lihe-solar}).  This is
not surprising due to the large primordial value from BBN\@.  Note, though,
that most \He{4} observations are made in extra-galactic \hii\ regions.  Thus
their evolution should not be expected to follow that of the solar
neighborhood.  We predict a linear relationship between \He{4} and O/H as
expected (figure~\ref{fig:Y-O}).  However the slope is much flatter than
appears in the observations as is a common failing in \chemevil\ models of the
type we have considered.

Deuterium and helium-3 are the two light elements that are most strongly
affected by \chemevil.  Thus \chemevil\ is essential to extract their
primordial abundances and test BBN\@.  The precise observation of D in the ISM
places very tight constraints on \chemevil\ models.  The evolution of \He{3}
in low mass stars also strongly affects the types of \chemevil\ models we can
construct to fit the observations.  As we noted above, the models we consider
here allow a range in the baryon-to-photon ratio, $\eta\approx
(3\hbox{--}5.5)\sci{-10}$.  Again this is due to the models we have chosen and
not a general requirement.  A wider range of starting values can produce
satisfactory fits to the observations (Olive \etal~1996).

The spread in both of these abundances is predicted to be quite small
(figure~\ref{fig:D-He-v-t}).  For the solar neighborhood this is in good
agreement with D observations along two different lines of sight in the ISM
(Linsky \etal~1993, 1995).  This lends some support to our assumption that
regions of size $\sim 10^5\Msun$ are well mixed in the solar neighborhood.
For \He{3} we would expect a larger spread based on observations in \hii\
regions (Balser \etal~1994) but again note that the results discussed here are
tuned for the solar neighborhood.  Notice that different histories do not
provide an explanation for the difference in D observations in quasar
absorption systems.  The observed difference is approximately an order of
magnitude.  Though it is premature to label either, let alone both, value as
primordial, it is even difficult to understand how they both can be
observations of D\@.  Starting from either value it is not known how the other
value could be reproduced (Jedamzik \& Fuller~1996) and why there is not a
distribution of values between these two extremes.

We must again point out that observations in \hii\ regions and in quasar
absorption systems are made in environments that can be quite different than
the solar neighborhood.  Thus we should not expect our models to reproduce
these regions.  Although the stochastic approach discussed here enjoys some
success in explaining the spread of abundances in the solar neighborhood, it
may be better suited for exploring quasar absorption systems and \hii\
regions.  These regions are more consistent with lower mass gas clouds in the
range $10^5\Msun$ as employed here.  Furthermore the observations of a scatter
of light element abundances in these regions may indicate the need for a
stochastic approach.  Quasar absorption systems may be ideal for this type of
approach since there are a number of different systems over a range of
redshifts observed.  Furthermore, metal lines have been observed in many of
these systems which provides constraints on the global features of the
\chemevil\ model.  Chemical evolution in quasar absorption systems has been
studied in the standard manner (Timmes, Lauroesch, \& Truran~1995; Malaney \&
Chaboyer~1996) and may benefit from the stochastic approach.

\section*{Acknowledgments}

I thank David Schramm and Michael Turner for their many suggestions during
this work and Rene Ong for many helpful comments on a draft of this
manuscript.  I gratefully acknowledge many useful discussions on \chemevil\
with Brian Fields, James Truran, Martin Lemoine, and Frank Timmes.  I also
thank Arnold Boothroyd and Corinne Charbonnel for many discussions of low mass
stellar yields and particularly Arnold Boothroyd for providing me with some of
his preliminary yield calculations.  Finally I thank Ken'ichi Nomoto for many
helpful discussions regarding both type II and type Ia supernovae.
Presented as a thesis to the Department of Physics, the University of Chicago,
in partial fulfillment of the requirements for the Ph.D. degree.
I thank the Institute for Nuclear Theory at the University of Washington for
its hospitality and the Department of Energy for partial support during the
completion of this work.
This work has been supported in part by the DOE (at Chicago and Fermilab) and
the NASA (at Fermilab through grant NAG 5-2788 and at Chicago through NAG
5-2770 and a GSRP fellowship) and by the NSF at Chicago through grant AST
92-17969.

\end{document}